\documentclass[12pt]{article}
\usepackage{fullpage}
\usepackage{latexsym,amssymb}

\newcommand{\qed}{\Box}

\newtheorem{lemma}{Lemma}

\newtheorem{proposition}[lemma]{Proposition}
\newtheorem{theorem}[lemma]{Theorem}

\newtheorem{corollary}[lemma]{Corollary}
\newtheorem{definition}[lemma]{Definition}

\newtheorem{claim}{\sc Claim}[lemma]

 \newcommand{\blackslug}{\hbox{\hskip 1pt \vrule width 4pt height 8pt
depth 1.5pt \hskip 1pt}}
 \newcommand{\QED}{\quad\blackslug\lower 8.5pt\null}
 \newcommand{\smQED}{\quad
  \vrule width1pt height6pt depth-1pt \kern -1pt
  \vrule width4pt height6pt depth-5pt \kern -4pt
  \vrule width4pt height2pt depth-1pt \kern -1pt
  \vrule width1pt height6pt depth-1pt }
 \newenvironment{proof}{{\bf Proof:}}{\QED}

\newcommand{\oldcomment}[1]{}
\newcommand{\junk}[1]{}

\renewcommand{\qed}{\hspace{1em}\hfill\rule{.5em}{.5em}}
\newcommand{\rubberskip}{\par\addvspace{\smallskipamount}}
\renewenvironment{proof}{\rubberskip\noindent{\bf Proof:}\ }{\qed
\rubberskip\noindent}

\newtheorem{freeaux}{}

\newenvironment{free}{\begin{freeaux}\hspace{-.9em}\rm}{\end{freeaux}}
\newcommand{\proofbox}{\rule{.5em}{.5em}}
\renewenvironment{proof}[1][Proof:]{\begin{free}{\bf {#1}}\ }{\ 
\hfill\proofbox\end{free}}

\newenvironment{subproof}{\begin{free}{\sc Proof:}\ }{\ \hfill$\Box$\end{free}}
\newpage
\newcommand{\eps}{\varepsilon}

\newcommand{\nestlab}{??}
\newcounter{algctr}\renewcommand{\thealgctr}{\Alph{algctr}}

\newlength{\lwidth}
\newlength{\lwidthi}
\newcounter{litemi}
\renewcommand{\thelitemi}{\nestlab.\arabic{litemi}}

\newlength{\lwidthii}
\newlength{\itemind}
\newlength{\itemindii}
\newcounter{litemii}
\renewcommand{\thelitemii}{\thelitemi.\arabic{litemii}}

\newlength{\lwidthiii}
\newlength{\itemindiii}
\newcounter{litemiii}
\renewcommand{\thelitemiii}{\thelitemii.\arabic{litemiii}}

\newlength{\lwidthiv}
\newlength{\itemindiv}
\newcounter{litemiv}
\renewcommand{\thelitemiv}{\thelitemiii.\arabic{litemiv}}

\newlength{\lwidthv}
\newlength{\itemindv}
\newcounter{litemv}
\renewcommand{\thelitemv}{\thelitemiv.\arabic{litemv}}

\newenvironment{prooft}[1]{\begin{free}{\bf Proof of Theorem~\ref{#1}:}\ }{\ 
\hfill\proofbox\end{free}}

\newcommand{\req}[1]{(\ref{#1})}
\newcommand{\idiv}{\hbox{\sc div}}
\newcommand{\mod}{\hbox{\sc mod}}

\newcommand{\drop}[1]{}

\newif\ifcom
\comtrue

\newcommand{\mtch}{}%{\marginpar{MT}}

\newcommand{\floor}[1]{\lfloor {#1} \rfloor}

\newcommand{\wordlength}{W}

\newcommand{\paren}[1]{\left({#1}\right)}

\title{Dynamic Ordered Sets with
Exponential Search Trees\footnote{This paper combines
results presented by the authors at the 37th FOCS 1996 \cite{And96}, 
the 32nd STOC 2000 \cite{AT00}, and
the 12th SODA 2001 \cite{AT01}}}
\author{Arne Andersson\\
Computing Science Department\\
Information Technology, Uppsala University\\
Box 311, SE - 751 05 Uppsala, Sweeden\\
\tt arnea@csd.uu.se http://www.csd.uu.se/$\sim$arnea
\and
Mikkel Thorup\\ 
AT\&T Labs--Research\\
Shannon Laboratory\\
180 Park Avenue, Florham Park\\
NJ 07932, USA\\
\tt mthorup@research.att.com }
\date{\ }

\begin{document}
\maketitle
\begin{abstract}
We introduce exponential search trees as a novel
technique for converting static polynomial space search structures 
for ordered sets into fully-dynamic linear space data structures.

This leads to an {\em optimal\/} bound of $O(\sqrt{\log n/\log\log n})$ for 
searching and updating a dynamic set of $n$ integer keys in
linear space. Here searching an integer $y$ means
finding the maximum key in the set which is smaller than or equal to $y$.
This problem is equivalent to the standard text book problem of maintaining an
ordered set (see, e.g., Cormen, Leiserson, Rivest, and Stein: 
{\em Introduction to Algorithms}, 2nd ed., MIT Press, 2001).

The best previous deterministic linear space bound was $O(\log
n/\log\log n)$ due Fredman and Willard from STOC 1990. No better
deterministic search bound was known using polynomial space.

We also get the following worst-case linear space 
trade-offs between the number $n$, the word length $\wordlength$, and the 
maximal key $U < 2^\wordlength$: $O(\min\{\log\log n+\log n/\log\wordlength,\ 
\log\log n\cdot\frac{\log \log U}{\log \log \log U}\})$. These trade-offs
are, however, not likely to be optimal.

Our results are generalized to finger searching and string searching, providing
optimal results for both in terms of $n$.
\end{abstract}
\drop{
\paragraph{Categories and Subject Descriptors} 
E.1 [Data] DATA STRUCTURES---{\em Lists stacks, and queues}, 
F.2.2 [Theory of Computation] ANALYSIS OF ALGORITHMS AND PROBLEM COMPLEXITY,
Nonnumerical Algorithms and Problems---{\em 
Computations on discrete structures}
G.2.2 [Mathematics of Computing] DISCRETE MATHEMATICS, Graph Theory---{\em 
Graph algorithms}
}

\section{Introduction}
\subsection{The Textbook Problem}
Maintaining a dynamic ordered set is a fundamental textbook problem. 
For example, following Cormen, Leiserson, Rivest, Stein: {\em Introduction to
Algorithms}~\cite[Part III]{CormAlgo}, the basic operations are:
\begin{description}
\item[Search $(S,k)$] Returns a pointer to an element in $S$ with key $k$, or return a null
pointer if $x$ is not in $S$.
\item[Insert $(S,x)$] Add $x$ to $S$, here $x$ is a pointer to an element.
The data structure may associate information with $x$,
e.g., parent and child pointers in binary search trees.
\item[Delete $(S,x)$] Remove $x$ from $S$, here $x$ is a pointer to an element in $S$.
\item[Predecessor/Successor $(S,x)$] Given that $x$ points at an element in $S$,
 return a pointer to
the next smaller/larger element in $S$ (or a null pointer if no such element exists).
\item[Minimum/Maximum $(S)$] Return a pointer to
the smallest/largest element in $S$ (or a null pointer if $S$ is empty).
\end{description}
To make the ordering total, we follow the convention
that if two elements have the same key, the last inserted element is larger.

For keys that can only be accessed via comparisons, all of the above
operations can be supported in $O(\log n)$ time\footnote{We
use the convention that logarithms are base 2 unless otherwise stated. Also,
$n$ is the number of stored elements}, which is best
possible.  

However, on computers, integers and floating point numbers are the
most common ordered data types. For such data types, represented as
lexicographically ordered, we can apply classical non-comparison based
techniques such as radix sort and hashing.  Historically, we note that
radix sort dates back at least to 1929 \cite{Com29} and hashing dates
back at least to 1956 \cite{Dum56}, whereas the focus on general
comparison based methods only dates back to 1959 \cite{FJ59}.

In this paper, we consider the above basic data types of integers and
floating point numbers. {\em Our main result is that we can support
all the above operations in $O(\sqrt{\log n/\log\log n})$ worst-case
time, and this common bound is best possible.}

The lower bound follows from a result of Beame and Fich~\cite{BF99}.
It shows that even if we just want to support Insert and Predecessor
in polynomial space, one of these two operations have a worst-case
bound of $\Omega(\sqrt{\log n/\log\log n})$, matching our common
upper bound. We note that one can find better bounds and trade-offs for
some of the individual operations. Indeed, we will
support Min, Max, Predecessor, Successor, and Delete in constant time, and
only do Insert and Search in $\Theta(\sqrt{\log n/\log\log n})$ time.

It is also worth noticing that if we just want to consider an 
incremental dictionary with Insert and Search, then our $O(\sqrt{\log
  n/\log\log n})$ Search time is the best known with $n^{o(1)}$ Insert
time.

\subsection{Extending the search operation}

In an ordered set, it is common to consider an extended version of Search:
\begin{description}
\item[Search $(S,k)$] Returns a pointer to an element in $S$ with the largest key
smaller than or equal to $k$, or null if $k$ is smaller than any key
in $S$.
\end{description}
Thus, if the key is not there, we do not just return a null pointer.
It is for this extended search that we provide our $O(\sqrt{\log
  n/\log\log n})$ upper bound. It is also for this search operation
that Beame and Fich~\cite{BF99} proved their $\Omega(\sqrt{\log
  n/\log\log n})$ lower bound, and that was even for a single extended
search in a static set $S$ for any given representation in polynomial
space. To see that this gives a lower bound just for Insert and
Predecessor, we can solve the static predecessor problem
as follows. First we insert all the elements of $S$ to create a
representation of $S$. The lower bound of Beame and Fich does not care
about the time for this preprocessing. To search $k$, we insert an
element with $k$ and ask for the predecessor of this element. The
lower bound of Beame and Fich then asserts that the Insert and
Predecessor operation together takes $\Omega(\sqrt{\log n/\log\log
  n})$ time in the worst-case, hence that at least one of the operations
has a worst-case lower bound of $\Omega(\sqrt{\log n/\log\log n})$.

In the rest of this paper, {\em search} refers to the extended version
whereas the primitive version, returning null if the key is not there,
is referred to as a {\em look-up}.

We will always maintain a sorted doubly-linked list with the stored elements
and a distinguished head and tail. With this list we support
Successor, Predecessor, Minimum, and Maximum in constant time.
Then Insert subsumes a Search identifying the element after which the
key is to be inserted. Similarly, if we want to delete by a key value,
rather than by a pointer to an element, a Search, or look-up, identifies an
element with the key to be deleted, if any. 

To isolate the search cost from the update cost, we talk about {\em finger
updates}, where for Insert, we are given the element after which the
key is to be inserted, and for Delete, we are given the element
to be deleted. Then an update with a key value is implemented
with a Search followed by a finger update. As we shall discuss later
in Section \ref{ssec:intro-finger}, we will implement all finger updates
in constant time. However, below, we mostly discuss common upper bounds
for searching and updating.

\subsection{History}

At STOC'90, Fredman and Willard \cite{FW93} surpassed the
comparison-based lower bounds for integer sorting and searching using
the features available in a standard imperative
programming languages such as C.  Their key result was an
$O(\log n/\log\log n)$ time bound for deterministic dynamic searching
in linear space. The time bounds for dynamic searching include both
searching and updates. Fredman and Willard's dynamic searching immediately
provided them an $O(n\log n/\log\log n)$ sorting routine. They asked the
fundamental question: {\em how fast can we search [integers on a
RAM]?\/} Since then, much effort has been spent on finding the
inherent complexity of fundamental searching and sorting problems.

In this paper, we introduce exponential search trees as a novel
technique for converting static polynomial space search structures for
ordered sets into fully-dynamic linear space data structures.  Based
on this we get an optimal worst-case bound of $O(\sqrt{\log n/\log\log
n})$ for dynamic integer searching. We note that this also provides the
best bounds for the simpler problem of membership and look-up queries
if one wants updates in $n^{o(1)}$ time  \cite{HMP01}.

Our results also extend to optimal finger search with constant
finger update, and to optimal string searching. 

\subsection{Model of computation} 
Our algorithms runs on a RAM, which models what we program in
imperative programming languages such as C. The memory is divided into
addressable words of length $\wordlength$. Addresses are themselves
contained in words, so $\wordlength\geq \log n$.  Moreover, we have a
constant number of registers, each with capacity for one word. The
basic instructions are: conditional jumps, direct and indirect
addressing for loading and storing words in registers, and some
computational instructions, such as comparisons, addition, and
multiplication, for manipulating words in registers.  The space
complexity is the maximal memory address used, and the time complexity
is the number of instructions performed. All keys are assumed to be
integers represented as binary strings, each fitting in one word.  One
important feature of the RAM is that it can use keys to compute
addresses, as opposed to comparison-based models of computation. This
feature is commonly used in practice, for example in bucket or radix
sort.

The restriction to integers is not as severe as it may seem.
Floating point numbers, for example, are ordered correctly, simply by
perceiving their bit-string representation as representing an integer.
Another example of the power of integer ordering is fractions of
two one-word integers. Here we get the right ordering if we carry 
out the division with floating point numbers with $2\wordlength$ bits 
of precession, and then just perceive the result as an integer. 
The above examples illustrate how integer
ordering can capture many seemingly different orderings that 
we would naturally be interested in.

The above word model is equivalent to a restriction that one only has
unit cost operations for integers that are polynomial in $n$ and the
integers in $X$. The later formulation goes back to Kirkpatrick and
Reisch in 1984 \cite{KR84}. We note that if we do not somehow limit
the size of the unit-cost integers, we get NP$=$P unless we start
ruling out common instructions such as multiplication and shifts.

\subsection{Historical developments}
As mentioned above, in 1990, Fredman and Willard \cite{FW93} showed that one 
can do dynamic searching in $O(\log n/\log\log n)$ time.  
They also showed that the
$O(\log n/\log\log n)$ bound could be replaced by an $O(\sqrt{\log
n})$ bound if we allow randomization or space unbounded in terms of
$n$. Fredman and Willard original construction was amortized, but
in 1992, Willard \cite[Lemma 3.3]{Wil00} showed that
the update bounds can be de-amortized so as to get worst-case bounds
for all operations.

In 1996, Andersson \cite{And96} introduced
{\em exponential search trees\/} as a general technique reducing
the problem of searching a dynamic set in linear space 
to the problem of creating a search structure for a static
set in polynomial time and space. The search time for the
static set essentially becomes the amortized search time in
the dynamic set. From Fredman and Willard \cite{FW93}, he got
a static search structure with $O(\sqrt{\log n})$ search time, and thus
he obtained an $O(\sqrt{\log n})$ time bound 
for dynamic searching in linear space.

In 1999, Beame and Fich showed that $\Theta(\sqrt{\log n/\log\log n})$ is
the exact complexity of searching a static set using polynomial space
\cite{BF99}.
Using the above mentioned exponential search trees, this gave
them $\Theta(\sqrt{\log n/\log\log n})$ amortized cost
for dynamic searching in linear space.

Finally, in 2000, Andersson and Thorup \cite{AT00} developed
a worst-case version of exponential search trees, giving
an {\em optimal\/}  $O(\sqrt{\log n/\log\log n})$ worst-case
time bound for dynamic searching. This is the main result presented
in this paper.

\subsection{Bounds in terms of the word length and the maximal key} 
Besides the
above mentioned bounds in terms of $n$, we get 
the following worst-case linear space 
trade-offs between the number $n$, the word length $\wordlength$, and the 
maximal key $U < 2^\wordlength$: $O(\min\{\log\log n+\log n/\log\wordlength,\ 
\log\log n\cdot\frac{\log \log U}{\log \log \log U}\})$.
The last bound should be compared with van Emde Boas' 
bound of $O(\log\log U)$ \cite{vB77,vBKZ77} that requires
randomization (hashing) in order to achieve linear space~\cite{MN90}. We note
that these bounds are probably not optimal. The best
lower bound on searching in terms of $U$ is Beame and Fich's 
$\Omega(\frac{\log \log U}{\log \log \log U})$ for the static case. 

\subsection{AC$^0$ operations}
As an additional challenge, Fredman and Willard \cite{FW93} asked how
quickly we can search on a RAM if all the computational instructions
are AC$^0$ operations.  A computational instruction is an AC$^0$
operation if it is computable by an $\wordlength^{O(1)}$-sized constant
depth circuit with $O(\wordlength)$ input and output bits. In the circuit
we may have negation, and-gates, and or-gates with unbounded fan-in.
Addition, shift, and bit-wise boolean operations are all AC$^0$
operations. On the other hand, multiplication is not. Fredman and Willard's
own techniques \cite{FW93} were heavily based on multiplication, but, as 
shown in \cite{AMT99} they can be implemented with AC$^0$ operations if
we allow some non-standard operations that are not part of the usual
instruction set. However, here we are interested in algorithms
using only standard operations so that they can be implemented in
a standard programming language such as C.

Concerning searching, our $O(\sqrt{\log n/\log\log n})$ search
structure is strongly based on multiplication. So far, even if we
allow amortization and randomization, no search structure using
standard AC$^0$ operations has been presented using polynomial space 
and $o(\log n)$ time, not even
for the static case. Without requirements of polynomial space,
Andersson \cite{And95} has presented a deterministic worst-case bound
of $O(\sqrt{\log n})$.  In this paper, we will present a linear space
worst-case AC$^0$ bound of $O((\log n)^{3/4+o(1)})$, thus surpassing the
$O(\log n)$ bound even in this restricted case.

\subsection{Finger searching}\label{ssec:intro-finger}
By {\em finger search} we mean that we can have a ``finger'' pointing
at a stored key $x$ when searching for a key $y$. Here a finger is
just a reference returned to the user when $x$ is inserted or searched
for. The goal
is to do better if the number $q$ of stored keys between $x$ and $y$ is
small. Also, we have {\em finger updates}, where for deletions, one
has a finger on the key to be deleted, and for insertions, one
has a finger to the key after which the new key is to be inserted.

In the comparison-based model of computation Dietz and Raman
\cite{DR94} have provided optimal bounds, supporting finger searches
in $O(\log q)$ time while supporting finger updates in constant time.
Very recently, Brodal et al.~\cite{BrodFing} have managed to match these
results on a pointer machine.

In this paper we present optimal bounds on the RAM; namely
$O(\sqrt{\log q/\log\log q})$ for finger search with constant time finger
updates. Also, we present the first finger search bounds that are efficient
in terms of the absolute distance $|y-x|$ between $x$ and $y$.

\subsection{String searching}
We will also consider the case of string searching where each key may
be distributed over multiple words.  Strings are then ordered
lexicographically. One may instead be interested in variable length
multiple-word integers where integers with more words are considered
larger. However, by prefixing each integer with its length, we reduce
this case to lexicographic string searching.

Generalizing search data structures for string searching is nontrivial
even in the simpler comparison-based setting.  The first efficient
solution was presented by Mehlhorn~\cite[\S III]{Meh84}.  While the
classical method requires weight-balanced search structures, our
approach contains a direct reduction to any unweighted search
structure.  With inspiration from~\cite{And96,AT00,BF99,Meh84} we show
if the longest common prefix between a key $y$ and the stored keys has
$\ell$ words, we can search $y$ in $O(\ell+\sqrt{\log n/\log \log n})$
time, where $n$ is the current number of keys. Updates can be done
within the same time bound.  Assuming that we can address the stored
keys, our extra space bound is $O(n)$.

The above search bound is optimal, for consider an instance of the 1-word
dynamic search problem, and give all keys a common prefix of $\ell$ words.
To complete a search we both need to check the prefix in $O(\ell)$ time,
and to perform the 1-word search, which takes 
$\Omega(\ell+\sqrt{\log n/\log \log n})$ \cite{BF99}.

Note that one may think of the strings as divided into characters
much smaller than words. However, if we only deal with one such
character at the time, we are not exploiting the full power
of the computer at hand.

\subsection{Techniques and main theorems}
Our main technical contribution is to introduce exponential search
trees providing a general reduction from the problem of maintaining
a worst-case dynamic linear spaced structure to the simpler problem of
constructing static search structure in polynomial time and
space. For example, the polynomial construction time allows us to 
construct a dictionary deterministically with look-ups in constant time.
Thus we can avoid the use of randomized hashing in, e.g., a van Emde Boas's
style data structure \cite{vB77,FW93,MN90}.
The reduction is captured by the following theorem:

\begin{theorem}\label{thm:search-red} 
Suppose a static search structure on
$d$ integer keys can be constructed in $O(d^{k-1})$, $k\geq 2$,  
time and space so that it
supports searches in $S(d)$ time. We can then construct
a dynamic linear space search structure that with $n$ integer keys supports 
insert, delete, and
searches in time $T(n)$ where 
\begin{equation}\label{eq:rec}
T(n)\leq T(n^{1-1/k})+O(S(n)).
\end{equation}
The reduction itself uses only standard
AC$^0$ operations.
\end{theorem}
We then prove the following result on static data structures:
\begin{theorem}\label{thm:static-search} 
In polynomial time and space, we can construct a deterministic data structure
over $d$ keys supporting searches in 
$O(\min\{\sqrt{\log d},\log \log U,
1  + \frac{\log d}{\log \wordlength}\})$ time 
where $\wordlength$ is the word length, and $U < 2^{\wordlength}$ is an
upper bound on the largest key. If we restrict ourselves to 
standard AC$^0$ operations, we
can support searches in $O((\log d)^{3/4+o(1)})$ worst-case time
per operation.
\end{theorem}
Above, the $\sqrt{\log d}$ and $\log\log U$ bounds were recently improved:
\begin{theorem}[Beame and Fich {\cite{BF99}}] \label{thm:BF99}
In polynomial time and space, we can construct a deterministic data structure
over $d$ keys supporting searches in 
$O(\min\{\sqrt{\log d/\log\log d},\frac{\log \log U}{\log \log \log U}\})$ 
time.
\end{theorem}
Applying the recursion from Theorem \ref{thm:search-red}, substituting $S(d)$
with (i) the two bounds in Theorem \ref{thm:BF99}, (ii) the last bound in the
min-expression in
Theorem \ref{thm:static-search}, and (iii) the AC$^0$ bound from 
Theorem \ref{thm:static-search}, we immediately
get the following four bounds:
\begin{corollary}\label{cor:search} 
There is a fully-dynamic deterministic linear space search structure
supporting insert, delete, and searches in worst-case time
\begin{equation}\label{eq:cor:search}
O\left(\min\left\{
\begin{array}{l} 
\sqrt{\log n/\log\log n} 
\\
\log\log n\cdot\frac{\log \log U}{\log \log \log U}
\\ 
\log\log n  + \frac{\log n}{\log \wordlength} 
\end{array}
\right\}\right)
\end{equation}
where $\wordlength$ is the word length, and $U < 2^{\wordlength}$ is
an upper bound on the largest key.
If we restrict ourselves to standard AC$^0$ operations, we
can support all operations in $O((\log n)^{3/4+o(1)})$ worst-case time
per operation.
\end{corollary}
It follows from the lower bound by
Beame and Fich \cite{BF99} that
our $O(\sqrt{\log n/\log\log n})$ bound
is optimal.

\subsubsection{Finger search}
A finger search version of Theorem \ref{thm:search-red} leads us
to the following finger search version of Corollary \ref{cor:search}: 
\begin{theorem}\label{thm:finger}
There is a fully-dynamic deterministic linear space search structure that
supports finger updates in constant time, and
given a finger to a stored key $x$, searches
a key $y>x$ in time
$$
O\left(\min\left\{
\begin{array}{l} 
\sqrt{\log q/\log\log q} 
\\
\log \log q\cdot \frac{\log \log (y-x)}{\log\log\log (y-x)}
\\ 
\log\log q + \frac{\log q}{\log \wordlength} 
\end{array}
\right\}\right)$$
where $q$ is the number of stored keys between $x$ and $y$. If we
restrict ourselves to AC$^0$ operations, we still get a bound
of $O((\log q)^{3/4+o(1)})$.
\end{theorem}

\subsubsection{String searching} We also present a general reduction
from string searching to 1-word searching:

\begin{theorem}\label{thm:string} For the dynamic string searching problem, if the longest
  common prefix between a key $x$ and the stored keys has $\ell$
  words, we can insert, delete, and search $x$ 
  in~$O(\ell+\sqrt{\log n/\log \log n})$
  time, where $n$ is the current number of keys. In addition to the
  stored keys themselves, our space bound is $O(n)$.
\end{theorem}

\subsection{Contents}
First, in Section \ref{sec:amortized}, we present a
simple amortized version of exponential search trees, and then
we de-amortize them in Section \ref{sec:worst-case}. In Section
\ref{sec:static} we construct the static search structures to
be used in the exponential search tree.
In Section \ref{sec:finger}, we describe the data structure
for finger searching. In Section \ref{sec:string}, we describe 
the data structure for string searching. In Section \ref{sec:other-appl}
we give examples of how the techniques of this paper have been applied
in other work. Finally, in Section \ref{sec:open-problem}, we finish
with an open problem.

\section{The main ideas and concepts in an amortized setting}\label{sec:amortized}
Before presenting our worst-case exponential search trees, we here
present a simpler amortized version from \cite{And96}, converting
static data structures into fully-dynamic amortized search structures.
The basic definitions and concepts of the amortized construction will
be assumed for the more technical worst-case construction.

An {\em exponential search
tree\/} is a leaf-oriented multiway search tree where the degrees of
the nodes decrease doubly-exponentially down the tree.  By {\em
leaf-oriented}, we mean that all keys are stored in the leaves of the
tree. Moreover, with each node, we store a {\em splitter\/}
for navigation: if a key arrives at a node, a local search among
the splitters of the children determines which child it belongs under.
Thus, if a child $v$ has splitter $s$ and its successor has splitter
$s'$, a key $y$ belongs under $v$ if $x\in [s,s')$. We require
that the splitter of an internal node equals the splitter
of its leftmost child.

We also maintain a doubly-linked list over the stored keys, providing
successor and predecessor pointers as well as maximum and minimum. A
search in an exponential search tree may bring us to the successor
of the desired key, but if the found key is to big, we just return
its predecessor.

In our exponential search trees, the local search at each internal node 
is performed using a static local search 
structure, called an {\em $S$-structure}. 
We assume that an $S$-structure over $d$ keys can be built 
in $O(d^{k-1})$ time and space and that it supports searches 
in $S(d)$ time. We define an exponential search tree over $n$ keys
recursively:
\begin{itemize}
\item The root has degree $\Theta(n^{1/k})$.
\item The splitters of the children of the root are stored in 
      a local $S$-structure with the properties stated above. 
\item The subtrees are exponential search trees over
      $\Theta(n^{1-1/k})$ keys.
\end{itemize}
It immediately follows that searches are supported in time
$T(n) = O\paren{S\paren{O(n^{1/k})}} + T\paren{O(n^{1-1/k})}$, which
is essentially the time bound we are aiming at.

An exponential search tree
over $n$ keys takes linear space. The space of 
the $S$-structure at a node of degree $d$ 
is $O\paren{d^{k-1}}$, and the total space $C(n)$ 
is essentially given by
\begin{eqnarray}
C(n) &=& O((n^{1/k})^{k-1}) +  n^{1/k}\cdot C(n^{1-1/k}) \nonumber \\
 \Rightarrow C(n) &=& O(n). \nonumber
\end{eqnarray}
The above calculation in not complete since we only recurse to
subproblems of size $O(n^{1-1/k})$, and then the direct solution is
not linear.  However, so far we are only sketching a simple amortized
version in order to introduce the general ideas. A rigorous argument
will be given for the real worst-case version (c.f. Lemma
\ref{lem:linear}).

Since $O(d^{k-1})$ bounds not only the space but also the construction time
for the $S$-structure at a degree $d$ node, the same argument gives
that we can construct an exponential
search tree over $n$ keys in linear time.

Recall that an update is implemented as a search, as described above,
followed by a finger update. A finger delete essentially just removes
the leaf with the element. However, if the leaf is the first child of
its parent, its splitter has to be transfered to its successor.  For a
finger insert of an element $u$ with key $k$, we get the element $v$
after which the key is to be inserted. We then also have to consider
the successor $v'$ of $v$. Let $s$ be the splitter of $v'$. If $k<s$,
we place $u$ after $v$ under the parent of $v$, and give $u$ splitter
$k$. If $k\leq s$, we place $u$ before $v'$ under the parent of $v'$, 
and give $u$ splitter $s$ and $v'$ its own key as splitter.

Balance is maintained in a standard fashion by global and partial
rebuilding. By the weight, $|t|$, of a (sub-)tree $t$ we mean the
number of leaves in $t$. By the weight, $|v|$, of a node $v$, we mean
the weight of the tree rooted at $v$.  When a subtree gets too heavy,
by a factor of 2, we split it in two, and if it gets too light, by a
factor of 2, we join it with its neighbor. Constructing a new subtree
rooted at the node $v$ takes $O(|v|)$ time.  In addition, we need to
update the $S$-structure at $v$'s parent $v$, in order to reflect the
adding or removing of a key in $u$'s list of child splitters.  Since
$v$ has $\Theta(|v|^{1/k})$ children, the construction time for $v$'s
$S$-structure is $O((|v|^{1/k})^{k-1})=O(|v|^{1-1/k})$. By definition,
this time is $O(|t|)$.  We conclude that we can reconstruct the
subtrees and update the parent's $S$-structure in time linear in the
weight of the subtrees.

Exceeding the weight constraints requires that a constant fraction of
the keys in a subtree have been inserted and deleted since the subtree
was constructed with proper weight. Thus, the reconstruction cost is
an amortized constant per key inserted or deleted from a tree. Since
the depth of an exponential search tree is $O(\log\log n)$, the update
cost, excluding the search cost for finding out were to update, is
$O(\log\log n)$ amortized. This completes our sketchy description of
amortized exponential search trees.

\section{Worst-case exponential search trees}\label{sec:worst-case}
The goal of this section is to prove the statement of 
Theorem~\ref{thm:search-red}:
\begin{quote}\it
Suppose a static search structure on
$d$ integer keys can be constructed in $O(d^{k-1})$, $k\geq 2$,  
time and space so that it
supports searches in $S(d)$ time. We can then construct
a dynamic linear space search structure that with $n$ integer keys supports 
insert, delete, and
searches in time $T(n)$ where 
$T(n)\leq T(n^{1-1/k})+O(S(n))$. The reduction itself uses only standard
AC$^0$ operations.
\end{quote}
In order to get from the amortized bounds above to worst-case bounds,
we need a new type of data structure.
Instead of a data structure where we occasionally rebuild entire subtrees,
we need a multiway tree which is something more in the style of a 
standard B-tree, where
balance is maintained by locally joining and splitting nodes.
By locally we mean that the joining and splitting is done just by
joining and splitting the children sequences.
This type of data structure is for example used by Willard \cite{Wil00}
to obtain a worst-case version of fusion trees.

One problem with our current definition of exponential search trees is
that the criteria for when subtrees are too large or too small depend
on their parents. If two subtrees are joined, the resulting subtree is
larger, and according to our recursive definition, this may imply that
all of the children simultaneously become too small, so they have to
be joined, etc. To avoid such cascading effects of joins and
splits, we redefine the exponential search tree as follows:
\begin{definition}
\label{def:exp}
In an exponential search tree all leaves are on the same
depth, and we define the height or level of a node to be the unique
distance from the node to the leaves descending from it. For a non-root
node $v$ at height $i>0$, the weight (number of descending leaves) is 
$|v|=\Theta(n_i)$ where $n_i=\alpha^{(1+1/(k-1))^{i}}$ and $\alpha=\Theta(1)$.
If the root has height $h$, its weight is $O(n_h)$.
\end{definition}
With the exception of the root, Definition \ref{def:exp} 
follows our previous definition
of exponential search trees, that is, if $v$ is a non-root node, it has
$\Theta(|v|^{1/k})$ children, each of weight $\Theta(|v|^{1-1/k})$.

Our main challenge is now to rebuild $S$-structures in the background
so that they remain sufficiently updated as nodes get joined and
split.  In principle, this is a standard task (see e.g. \cite{WL85}).
Yet it is a highly delicate puzzle which is typically either not done
(e.g. Fredman and Willard  \cite{FW93} only claimed 
amortized bounds for their original fusion trees), 
or done with rather incomplete sketches (e.g. Willard \cite{Wil00} 
only presents a 
2-page sketch of his
de-amortization of fusion trees). Furthermore, our
exponential search trees pose a new complication; namely that when we
join or split, we have to rebuild, not only the $S$-structures of the
nodes being joined or split, but also the $S$-structures of their parent.
For contrast, when Willard \cite{Wil00} de-amortizes fusion trees,
he actually uses the ``atomic heaps'' from \cite{FW94} as $S$-structures,
and these atomic heaps support insert and delete in constant time. 
Hence, when nodes get joined or split, he can just delete or insert 
the splitter between them directly in the parents $S$-structure, 
without having to rebuild it.

In this section, we will present a general 
quotable theorem about rebuilding, thus making proper de-amortization much
easier for future authors. 

\subsection{Join and split with postprocessing}\label{ssec:balance}
As mentioned, we are going to deal generally with multiway trees where joins
and splits cannot be completed in constant time. For the moment,
our trees are only described structurally with a children list for each 
non-leaf node. Then joins and splits can be done in constant time. However,
after each join or split, we want to allow for some unspecified
{\em postprocessing\/} before the involved nodes can participate in new 
joins and splits. This postprocessing time will, for example, be used to 
update parent pointers and $S$-structures.

The key issue is to schedule the postprocessing, possibly involving
reconstruction of static data structures, so that we obtain good
worst-case bounds. We do this by dividing each postprocessing into
{\em local update steps} and ensuring that each update only uses a few
local update steps at the same time as each postprocessing is given enough 
steps to complete. The schedule is independent of how these update
steps are performed.

To be more specific in our structural description of a tree, 
let $u$ be the predecessor of $v$ in their
parent's children list $C$. A {\em join\/} of $u$ and
$v$ means that we append the children list of $v$ to that of $u$ so
that $u$ adopts all these children from $v$. Also, we delete $v$ from
$C$. Similarly, a {\em split\/} of a node $u$ at its child $w$ means
that we add a new node $v$ after $u$ in the children list $C$ of $u$'s
parent, that we cut the children list of $u$ just before $w$, and make
the last part the children list of $v$. Structural splits and joins
both take constant time and are viewed as atomic operations. In the
postprocessing of a join, the resulting node is not allowed to
participate in any joins or splits. In the postprocessing of a split, the
resulting nodes are neither allowed to participate directly in a join
or split, nor is the parent allowed to split between them.

We are now in the position to present our general theorem on 
worst-case bounds for joining and
splitting with postprocessing:

\begin{theorem}\label{thm:balance} 
Given a number series $n_1, n_2, \ldots$, with $n_1\geq 84$, $n_{i+1} >
18 n_i$, we can schedule split and joins to maintain a multiway tree where
each non-root node $v$ on height $i>0$ has weight between $n_i/4$ and $n_i$.
A root node on height $h>0$ has weight at most $n_h$ and at least
$2$ children. The schedule gives the following properties:

(i) When a leaf $v$ is inserted or deleted, for each node $u$ on the
path from $v$ to the root 
the schedule use one local update step contributing
to the postprocessing of at most one join or split involving 
either $u$ or a neighbor of $u$. 

(ii) For each split or join at level $i$
the schedule ensures that we have $n_i/84$ local update steps 
available for postprocessing, including one at the time of
the split or join.

(iii) If the time of a local update step on level
$i$ is bounded by $t_i=\Omega(1)$, each update is supported in
$O(\sum_{i=1}^h t_i)$ time.
\end{theorem}
In our exponential search tree, we will have $t_i=O(1)$, but $t_i=\omega(1)$
has been useful in connection with priority queues \cite{AT00}.

The above numbers ensure that a node which is neither root nor leaf
has at least $(n_i/4)/n_{i-1}=18/4>4$ children. If the
root node is split, a new parent root is generated implicitly.
Conversely, if the root's children join to a single child, the root is
deleted and the single child becomes the new root. The proof
of Theorem \ref{thm:balance} is rather delicate, and deferred till
later. Below we show how to apply Theorem \ref{thm:balance} in
exponential search trees. As a first simple application of the schedule,
we show how to maintain parents.
\begin{lemma}\label{lem:parents} In Theorem \ref{thm:balance}, 
the parent of any node can be computed in constant time. 
\end{lemma}
\begin{proof}
With each node, we maintain a parent pointer, which points to
the true parent, except, possibly during the postprocessing
of a split or join. Split and joins are handled equivalently. 
Consider the case of a join of $u$ and $v$
into $v$. During the postprocessing, we will redirect all
the parent pointers of the old children of $v$ to point to
$u$. Meanwhile, we will have a forward pointer from $v$ to $u$ so
that parent queries from any of these children can be answered
in constant time, even if the child still points to $v$.

Suppose that the join is on level $i$. Then $v$ could not have more
than $n_i$ children. Hence, if we redirect $84$ of their parent
pointers in each local update step, we will be done the end of the
postprocessing of Theorem \ref{thm:balance}. The redirections are done
in a traversal of the children list, starting from the old first child
of $v$. One technical detail is, however, that we may have join and
split in the children sequence. Joins are not a problem, but for split
we make the rule that a if we split $u'$ into $u'$ and $v'$, $v'$
inherits the parent pointer of $v'$. Also, we make the split rule that
if the parent pointer of $u'$ is to a node like $v$ with a forward
pointer to a node like $u$ that $v$ is being joined to, we redirect
the next child in the above redirection traversal from $v$. This way
we make sure that the traversal of the old children list of $v$
is not delayed by splits in the list.
\end{proof}
For our exponential search trees, we will use the postprocessing for
rebuilding $S$-structures. We will still keep a high level of
generality to facilitate other applications, such as, for example, 
a worst-case version of the original fusion trees \cite{FW93}.
\begin{corollary}\label{cor:search-balance} 
Given a number series $n_0,n_1, n_2, \ldots$, with $n_0=1$, $n_1\geq 84$, 
$n_{i+1} >
18 n_i$, we maintain a multiway tree where each node on height $i$ 
which is neither the root nor a leaf node has weight between $n_i/4$ and 
$n_i$. If an $S$-structure for a node on height $i$ 
can be built in $O(n_{i-1}t_i)$ time, $t_i=\Omega(1)$, 
we can maintain $S$-structures
for the whole tree in $O(\sum_{i=1}^h t_i)$ time per finger update.
\end{corollary}
\begin{proof}
In Section \ref{sec:amortized}, we described how
a finger update, in constant time, translates into
the insertion or deletion of a leaf. We can
then apply Theorem \ref{thm:balance}.

Our basic idea is that we have an ongoing periodic rebuilding of the
$S$-structure at each node $v$. A period starts by scanning the splitter
list of the children of $v$ in $O(n_i/n_{i-1})$ time. It then creates a new
$S$-structure in $O(n_{i-1}t_i)$ time, and finally, in constant time, it
replaces the old $S$-structure with the new $S$-structure.  The whole
rebuilding is divided into $n_{i-1}/160$ steps, each taking $O(t_i)$ time.

Now, every time an update contributes to a join or split postprocessing
on level $i-1$, we
perform one step in the rebuilding of the $S$-structure of the parent $p$, 
which is on level $i$. 
Then Theorem \ref{thm:balance} ascertains that we perform 
$n_{i-1}/84$ steps on $S(p)$ during the postprocessing, and hence we
have at least one complete rebuilding of $S(p)$ with(out) the splitter
created (removed) by the split (join).

When two neighboring nodes $u$ and $v$ on level $i-1$ join,
the next rebuilding of $S(u)$ will automatically include the old
children of $v$. The rebuilding of $S(v)$ is continued for all updates
belonging under the old $v$ until the $S(u)$ is ready to take over, but
these updates will also promote the rebuilding of $S(u)$. This way we
make sure that the children of $v$ and $u$ do not experience any delay
in the rebuilding of their parents $S$-structure. Note that $S(u)$
is completely rebuilt in $n_{i-2}/84$ steps which is much less than
the $n_{i-1}$ steps we have available for the postprocessing.

During the postprocessing of the join, we may, internally, have to
forward keys between $u$ and $v$. More precisely, if a key arrives at
$v$ from the parents $S$-structure and $S(u)$ has been updated to take
over $S(v)$, the key is sent through $S(u)$. Conversely, $S(v)$ is
still in use; if a key arriving at $u$ is larger than or equal to the
splitter of $v$ it is sent through $S(v)$.

The split is implemented using the same ideas: all updates for the two
new neighboring nodes $u$ and $v$ promote both $S(u)$ and $S(v)$.
For $S(u)$, we finish the current rebuilding over all the children 
before doing a rebuild excluding the children going to $v$. By the end
of the latter rebuild, $S(v)$ will also have been completed. If a key
arrives at $v$ and $S(v)$ is not ready, we send it through
$S(u)$. Conversely, if $S(v)$ is ready and a key arriving at $u$ is
larger than or equal to the splitter of $v$, the key is sent though
$S(v)$.
\end{proof}
Below we establish some simple technical lemmas verifying that
Corollary \ref{cor:search-balance} applies
to the exponential search trees from Definition \ref{def:exp}.
The first lemma shows that the number sequences $n_i$ match.
\begin{lemma}\label{lem:exp-bound}  
With $\alpha=\max\{84^{(k-1)/k},18^{(k-1)^2/k}\}$ and
$n_i=\alpha^{(1+1/(k-1))^i}$ as in Definition \ref{def:exp},
$n_1\geq 84$ and $n_{i+1}/n_i\geq 18$ for $i\geq 1$.
\end{lemma}
\begin{proof} $n_1\geq (84^{(k-1)/k})^{1+1/(k-1)}=84$ and
$n_{i+1}/n_i=n_{i+1}^{1/k}\geq n_{2}^{1/k}\geq \alpha^{(k/(k-1))^2/k}$.
\end{proof}
Next, we show that the $S$-structures are built fast enough.
\begin{lemma}\label{lem:simple-update} 
With Definition \ref{def:exp}, creating an $S$-structure 
for a node $u$ on level $i$ takes $O(n_{i-1})$ time, and the total
cost of a finger update is $O(\log\log n)$.
\end{lemma}
\begin{proof}
Since $u$ has degree at most $4n_i^{1k/}$,
the creation takes $O((n_i^{1/k})^{k-1})=O((n_i^{1-1/k})=O(n_{i-1})$ time.
Thus, we get $t_i=O(1)$ in Corollary \ref{cor:search-balance}, corresponding
to a finger update time of $O(\log\log n)$.
\end{proof}
Since $S(n)=\Omega(1)$, any time bound derived from Theorem
\ref{thm:search-red} is $\Omega(\log\log n)$, dominating our cost of a
finger update. 

Next we give a formal proof that the recursion formula of Theorem 
\ref{thm:search-red} holds.
\begin{lemma} 
\label{lem:search-red}
Assuming that the cost for searching in a node of degree~$d$ is
$O(S(d))$, the search time for an $n$ key exponential search tree
from Definition \ref{def:exp} is bounded 
by $T(n)\leq T(n^{1-1/k})+O(S(n))$ for $n=\omega(1)$.
\end{lemma}
\begin{proof} 
Since $n_{i-1}=n_i^{1-1/k}$ and since the degree of a
level $i$ node is at most $4n_i^{1/k}$, the search time starting
just below the root at level $h-1$ is bounded by $T'(n_{h-1})$ where
$n_{h-1}< n$ and $T'(m)\leq T'(m^{1-1/k})+O(S(4m^{1/k}))$. 
Moreover, for $m=\omega(1)$, $4m^{1/k}>m$, so $O(S(4m^{1/k}))=O(S(m))$.

The degree of the root is bounded by $n$, so the
search time of the root is at most $S(n)$. Hence our
total search time is bounded by $S(n)+T'(n_{h-1})
=O(T(n))$. Finally, the $O$ in $O(T(n))$ is superfluous because of
the $O$ in $O(S(n))$.
\end{proof}
Finally, recall that our original analysis, showing that exponential search
trees used linear space, was not complete. Below comes the formal proof.
\begin{lemma}\label{lem:linear} The exponential search trees from
Definition \ref{def:exp} use linear space.
\end{lemma}
\begin{proof} Consider a node $v$ at height $i$. The number of keys
below $v$ is at least $n_i/4$. Since $v$ has degree at most $4 n_i^{1/k}$, 
the space of the $S$-structure by $v$ is 
$O((4 n_i^{1/k})^{k-1})=O(n_i^{1-1/k})$. Distributing this space
on the keys descending from $v$, we get  $O(n_i^{-1/k})$ space per key.

Conversely, for a given key, the space attributed to the key by
its ancestors is $O(\sum_{i=0}^{h}n_i^{-1/k})=O(1)$.
\end{proof}
The above lemmas establish that {\bf Theorem \ref{thm:search-red} holds
if we can prove Theorem \ref{thm:balance}}.

\subsection{A game of weight balancing}\label{ssec:balance-game}
In order to prove Theorem \ref{thm:balance}, we consider
a game on lists of weights. In relation to Theorem \ref{thm:balance},
each list represents
the weights of the children of a node on some
fixed level. The purpose of the game is crystallize 
what is needed for balancing on each level.
Later, in a bottom-up induction, we will apply the game
to all levels.

First we consider only one list.

Our goal is to maintain balance, that is, for some parameter $b$,
called the ``latency'', all weights should be of size $\Theta(b)$. An
adversary can update an arbitrary weight, incrementing or decrement it
by one.  Every time the adversary updates a weight, we get to work
locally on balance. We may join neighboring weights $w_1$ and $w_2$
into one $w=w_1+w_2$, or split a weight $w$ into two $w_1$ and $w_2$,
$w_1+w_2=w$.

Each join or split takes $b$ steps. A join, replacing $w_1w_2$ with
$w=w_1+w_2$, takes place instantly, but requires a $b$ step postprocessing. 
A split, replacing $w$ with $w_1w_2$, $w_1+w_2=w$, happens
at a time chosen by the advisary during the $b$ steps. The adversary
must fulfill that $|w_1-w_2|\leq\Delta b$, where
$\Delta$ is called the ``split error''. This should be satisfied from the
split is done and until the split process are completed.

Every time a weight involved in a join or split process is updated,
we get to do one step on the process. However, we may also need to
work on joining or splitting of neighboring weights. More precisely, a
weight is said to be {\em free\/} if it is not involved in a split or join
process. In order to ensure that free weights do not get too small, we
need a way of requesting that they soon be involved in a join or
split process. To this end, we introduce tieing: if a free weight $v$
has an involved neighbor $w$, we may {\em tie\/} $v$ to $w$, and then each
update to $v$ will progress that process on $w$ by one step. 

Recall again that we are really working on a family of lists
which the adversary may cut and concatenate. We note that the adversary
may only do one operation at the time; either an update or a cut or a
split, and for each operation, we get time to respond with an update
step. Our only restriction on the adversary is that it is not allowed
to cut between weights involved in a join or split process, or cut off
a weight tied to such a process.

\begin{proposition}
\label{lem:buckets} Let $\mu>1$. Let $b$ be the ``latency'' , and 
$\Delta$ be the ``split error''. A list is ``neutral'' if
all weights are strictly between $(\mu+3)b$ and
$(2\mu+\Delta+9)b$. We start with neutral lists, and neutral lists
can be added and removed at any time. As long as all lists have a total weight
$>(\mu+3)b$, there is a protocol guaranteeing that the each weight is
between $\mu b$ and $(3\mu+\Delta+14)b$, and that the total weight of
any uncuttable segment of a list is at most $(5\mu+\Delta+19)b$.

In particular, for $\Delta=7$ and $\mu=21$, with start weights strictly 
between $24b$ and $58b$, we guarantee 
that the weights stay between $21b$ and $84b$ and that the maximum 
uncuttable segment is of size at most $131b$.
\end{proposition}
In our application, \mtch we define $B=84b$ so the base
segments of size between $\frac14 B$ and $B$, and the uncuttable
segments are of size below $2B$. A list is then neutral if all weights
are between $\frac{24}{84}B<\frac13 B$ and $\frac{58}{84}B>\frac23 B$.
We are then given $b=\frac1{84}B$ update steps to perform a join or split,
and during a split, the two new weights should differ by at most
$\frac1{12}B$.

As a first step in the proof of Proposition \ref{lem:buckets}, we present
the concrete protocol itself.
\begin{itemize}
\item[(a)] If a free weight gets up to $sb$, $s=2\mu+\Delta+9$, we 
split it. (Recall that we even allow an adversary to postpone the event
when the split is actually made.)
\item[(b)] If a free weight $v$ gets down to $mb$, $m=\mu+3$ and has a free 
neighbor $w$, we join $v$ and $w$, untieing $w$ from any 
other neighbor.
\item[(c)] If a free weight $v$ gets down to $mb$ and
has no free neighbors, we tie it to any one of its neighbors. If
$v$ later gets back up above $mb$, it is untied again.
\item[(d)] When we finish a join postprocessing, if the resulting weight is
$\geq sb$, we immediately split it. If a neighbor was tied
to the joined node, the tie is transfered to the nearest node resulting
from the split. 

If the weight $v$ resulting from the join is $<sb$ and $v$ is tied
by a neighbor, we join with the neighbor that tied $v$ first.
\item[(e)] At the end of a split postprocessing, if any of the
resulting nodes are tied by a neighbor, it joins with that
neighbor. Note here that since resulting nodes
are not tied to each other, there cannot be a conflict.
\end{itemize}
Note that our protocol is independent of legal cuts and concatenations
by the adversary, except in (c) which requires that a free weight
getting down to $(\mu+3)b$ has at least one neighbor. This is, however,
ensured by the condition from Proposition \ref{lem:buckets} that
each list has total weight strictly larger than $(\mu + 3) b$.
\begin{lemma}\label{lem:nodesize}\ 
\begin{itemize}
\item[(i)] Each free weight is between $\mu b$ and $sb=(2\mu+\Delta+9)b$.
\item[(ii)] The weight in a join process is between 
$(m+\mu-1)b=(2\mu+2)b>mb$
and $(s+m+1)b=(3\mu+\Delta+13)b>sb$.
\item[(iii)] In a split process, the total weight is at most $(3\mu+\Delta+14))b$ and the split weights are between
$((s-1-\Delta)/2) b=(\mu+4)b>mb$ and $((s+m+2+\Delta)/2) b
=(1\mu+\Delta+8)b<sb$.
\end{itemize}
\end{lemma}
\begin{proof} First we prove some simple claims.
\begin{claim}
If (i), (ii), and (iii) are true when a join process starts, 
then (ii) will remain satisfied for that join process.
\end{claim}
\begin{subproof}
For the upper bound note that when the join is started, none of
the involved weights can be above $sb$, for then we would have split
it. Also, a join has to be initiated by a weight of size at most $mb$, so
when we start the total weight is at most $(s+m)b$, and during
the postprocessing, it can increase by at most $b$.

For the lower bound, both weights have to be at least
$\mu b$. Also, the join is either initiated as in (b) by a weight
of size $mb$, or by a tied weight coming out from a join or split,
which by (ii) and (iii) is of size at least $mb$, so we start with a total
of at least $(\mu+m)b$, and we loose at most $b$ in the postprocessing.
\end{subproof}
\begin{claim}
If (i), (ii), and (iii) are true when a split process starts, 
then (iii) will remain satisfied for that split process.
\end{claim}
\begin{subproof}
For the lower bound, we know that a split is only initiated for a weight
of size at least $sb$. Also, during the process, we can loose at most
$b$, and since the maximal difference between the two weights is
$\Delta b$, the smaller is of size at least $(s-1-\Delta)/2 b$.

For the upper bound, the maximal weight we can start with is one coming
out from a join, which by (ii) is at most
$(s+m+1)b$. We can gain at most $b$ during the split processing, so
the larger weight is at most $((s+m+2+\Delta)/2)b$.
\end{subproof}

We will now complete the proof of Lemma~\ref{lem:nodesize}
by showing that there
cannot be a first violation of (i) given that (ii) and (iii) have not
already been violated. The only way we can possibly get a weight
above $sb$ is one coming out from a join as in (ii), but then
by (d) it is immediately split, so it doesn't become free.

To show by contradiction that we cannot get a weight below $\mu b$, 
let $w$ the first weight getting down
below $\mu b$ keys. When $w$ was originally created by
(ii)  and (iii), it was of size $>mb$, so
to get down to $\mu b$, there must have been a last time where it
got down to $m b$. It then tied itself to an involved neighboring weight
$w'$. If $w'$ is involved in a split, we know that when $w'$ is done, the
half nearest $w$ will immediately start joining with $w$ as in (e). 
However, if $w'$ is involved
in a join, when done, the resulting weight may start joining with a 
weight $w''$ on the
other side. In that case, however, $w$ is the first weight to tie to
the new join. Hence, when the new join is done, either $w$ starts joining
with the result, or the result get split and then $w$ will join with
the nearest weight coming out from the split. In the worst case, $w$ will
have to wait for two joins and one split to complete before it
gets joined, and hence it can loose at most $3b=(m-\mu)b$ while waiting
to get joined.
\end{proof}

\paragraph{Proof of Proposition \ref{lem:buckets}}
By Lemma \ref{lem:nodesize}, all weights
remain between $\mu b$ and $(3\mu+\Delta+13)b$.
Concerning the maximal size of an uncuttable segment,
the maximal total weight involved in split or join is
$(m+s+1)b$, and by (b)we can have a weight of size at most $mb$ tied
from either side, adding up to a total of
$(3m+s+1)b=(5\mu+\Delta+19)b$.

\subsection{Applying the protocol}\label{ssec:apply-prot}
We now want to apply our protocol in order to prove Theorem \ref{thm:balance}:
\begin{quote}\em
Given a number series $n_1, n_2, \ldots$, with $n_1\geq 84$, $n_{i+1} >
18 n_i$, we can schedule split and joins to maintain a multiway tree where
each non-root node $v$ on height $i>0$ has weight at between $n_i/4$ and $n_i$.
A root node on height $h>0$ has weight at most $n_h$ and at least
$2$ children. The schedule gives the following properties:

(i) When a leaf $v$ is inserted or deleted, for each node $u$ on the
path from $v$ to the root 
the schedule use one local update step contributing
to the postprocessing of at most one join or split involving 
either $u$ or a neighbor of $u$. 

(ii) For each split or join at level $i$
the schedule ensures that we have $n_i/84$ local update steps 
available for postprocessing.

(iii) If the time of a local update step on level
$i$ is bounded by $t_i=\Omega(1)$, each update is supported in
$O(\sum_{i=1}^h t_i)$ time.
\end{quote}
For each level $i<h$, the nodes are partitioned in children lists of nodes
on level $i+1$. We maintain these lists using the scheduling of Proposition
\ref{lem:buckets}, with latency $b_i=n_i/84$ and split error $\Delta=7$. With
$\mu=21$, this will give weights between $21 b$ and $84b$, as
required. We need to ensure that the children list of a node
on level $i$ can be cut so that the halves differ by at most
$\Delta b_i$. For $i=1$, this is trivial, in that the children list
is just a list of leaves that can be cut anywhere, that is,
we are OK even with $\Delta=1$. For $i>1$, inductively,
we may assume that we have the required difference of $\Delta b_{i-1}$ on
level below, and then, using Proposition \ref{lem:buckets}, we
can cut the list on level $i$ with a difference of $131 b_{i-1}$. However,
$b_i\geq 18 b_{i-1}$, so $131 b_{i-1}\leq 7 b_i$, as required.

Having dealt with each individual level, three unresolved problems remain:
\begin{itemize}
\item How do we for splits in constant time find a good place to cut
the children list?
\item How does the protocol apply as the height of the tree changes?
\item How do we actually find the nodes on the path from the leaf $v$ to the
root?
\end{itemize}
\paragraph{Splitting in constant time}
For each node $v$ on level $i$, our goal is to maintain a good cut child 
in the sense that when cutting at that child, the lists will not differ 
by more than $\Delta b_i$. We will always maintain the 
sum of the weights of
the children preceding the cut child, and comparing that with the weight
of $v$ tells us if it is at good balance. If an update makes the preceding 
weight to large, we move to the next possible cut child to the right,
and conversely, if it gets to small, we move the cut child to the left. 
A possible cut is always at most 4 children away, so the above
shifts only take constant time. Similarly, if the cut child stops being
cuttable, we move in the direction that gives us the best balance.      

When a new list is created by a join or a split, we need to find a new
good cut child. To our advantage, we know that we have at least
$b_i$ update steps before the cut child is needed. We can therefore
start by making the cut child the rightmost child, and every time
we receive an update step for the join, we move to the right, stopping
when we are in balance. Since the children list is of length 
$O(n_i/n_{i-1})$,  we only need to move a constant number of children to 
the right in each update step in order to ensure balance before the 
postprocessing is ended.

\paragraph{Changing the height}
A minimal tree has a root on height $1$, possibly with $0$
children.  If the root is on height $h$, we only apply the 
protocol when it has weight at least $21b_h$, splitting
it when the protocol tells us to do so. Note that there is no
cascading effect, for before the split, the root
has weight at most $84b_h$, and this is the weight 
of the new root at height $h+1$. However $b_h\leq b_{h+1}/18$, so it 
will take many updates before the new root reaches the 
weight~$21 b_{i+1}$. The $S$-structure and pointers of the new root
are created during the postprocessing of the split of the old
root. Conversely, we only loose a root at height $h+1$ when
it has two children that get joined into one child. The 
cleaning up after the old root, i.e.\ the removal of its $S$-structure
and a constant number of pointers, is done in the postprocessing
of the join of its children. We note that the new
root starts with weight at least $21 b_h$, so it has at least
$21 b_h/84 b_{h-1}\geq 18/4>4$ children. Hence it will survive
long enough to pay for its construction.

\paragraph{Finding the nodes on the path from the leaf $v$ to the root}
The obvious way to find the nodes on the path from the leaf $v$ to the
root is to use parent pointers, which according to Lemma
\ref{lem:parents} can be computed in constant time. Thus, we can prove
Theorem \ref{thm:balance} from Lemma \ref{lem:parents}. The only
problem is that we used the schedule of Theorem \ref{thm:balance} to
prove Lemma \ref{lem:parents}.  To break the circle, consider the
first time the statement of Theorem \ref{thm:balance} or of Lemma
\ref{lem:parents} is violated. If the first mistake is a mistaken
parent computation, then we know that the scheduling and weight
balancing of Theorem \ref{thm:balance} has not yet been violated, but
then our proof of Lemma \ref{lem:parents} based on Theorem
\ref{thm:balance} is valid, contradicting the mistaken parent
computation. Conversely, if the first mistake is in Theorem
\ref{thm:balance}, we know that all parents computed so far were
correct, hence that our proof of Theorem \ref{thm:balance} is
correct. Thus there cannot be a first mistake, so {\bf we conclude
that both Theorem \ref{thm:balance} and Lemma \ref{lem:parents} are
correct}.

%------------------------------------

\section{Static search structures}\label{sec:static}

In this section, we will prove Theorem~\ref{thm:static-search}:
\begin{quote}\it
In polynomial time and space, we can construct a deterministic data structure
over $d$ keys supporting searches in 
$O(\min\{\sqrt{\log d},\log \log U,
1  + \frac{\log d}{\log \wordlength}\})$ time 
where $\wordlength$ is the word length, and $U < 2^{\wordlength}$ is an
upper bound on the largest key. If we restrict ourselves to 
standard AC$^0$ operations, we
can support searches in $O((\log d)^{3/4+o(1)})$ worst-case time
per operation.
\end{quote}
To get the final bounds in Corollary~\ref{cor:search}, 
we actually need to improve the first bound in the min-expression to
$O(\sqrt{\log n/\log\log n})$ and the second bound 
to $O(\log\log U/\log\log\log U)$.
However, the improvement is by Beame and Fich \cite{BF99}. 
We present our bounds here
because (i) they are simpler (ii) the improvement by Beame and Fich
is based on our results.

\subsection{An improvement of fusion trees}

\noindent
Using our terminology,
the central part of the fusion tree is a 
static data structure with the following properties:

\begin{lemma} (Fredman and Willard)
\label{lem:FW}
For any $d$, $d = O\paren{\wordlength^{1/6}}$, 
A static data structure containing $d$ keys
can be constructed in $O\paren{d^4}$ time and space, 
such that it supports neighbor queries 
in~$O(1)$ worst-case time.
\end{lemma}

Fredman and Willard used this
static data structure to implement a B-tree 
where only the upper levels
in the tree contain B-tree nodes, all having the same degree (within 
a constant factor).
At the lower levels, traditional (i.e. comparison-based)
weight-balanced trees were used.
The amortized cost of searches and updates
is $O(\log n/\log d +\log d)$
for any $d = O\paren{\wordlength^{1/6}}$.
The first term corresponds to the
number of B-tree levels and the
second term corresponds to the height of the weight-balanced trees.

Using an exponential search tree instead of the Fredman/Willard structure, 
we avoid the
need for weight-balanced trees at the bottom at the same
time as we improve the complexity for large word sizes.

\begin{lemma}
\label{lem:FWgen}
A static data structure containing $d$ keys
can be constructed in $O\paren{d^4}$ time and space, 
such that it supports neighbor queries 
in~$O\paren{\frac{\log d}{\log \wordlength}+1}$ worst-case time.
\end{lemma}

\begin{proof}
We just construct a static B-tree where each
node has the largest possible degree according to Lemma \ref{lem:FW}.
That is, it has a degree of $\min\paren{d,\wordlength^{1/6}}$.
This tree satisfies the conditions of the lemma.
\end{proof}

\begin{corollary}
\label{cor:FWpart}
There is a data structure occupying linear
space for which the
worst-case cost of a search and update is
$O\paren{\frac{\log n}{\log \wordlength} + \log\log n}$
\end{corollary}

\begin{proof}
Let $T(n)$ be the worst-case cost.
Combining Theorem \ref{thm:search-red} and 
Lemma~\ref{lem:FWgen} gives that
$$
T(n) = O\paren{\frac{\log n}{\log \wordlength} + 1 + T\paren{n^{4/5}}}.
$$
\end{proof}

\subsection{Tries and perfect hashing}
\label{sec:trie}

\noindent
In a binary trie, a node at depth $i$
corresponds to an $i$-bit prefix of one (or more) of the keys stored in the trie. 
Suppose we could access a node by its prefix in constant time by means of a
hash table, i.e. without traversing the path down to the node.
Then, we could find a key~$x$, or $x$'s nearest neighbor, 
in~$O(\log \wordlength)$ time by a binary search
for the node corresponding to~$x$'s longest matching prefix.
At each step of the binary search, we look in the hash table
for the node corresponding to
a prefix of $x$; if the node is there we try with a longer prefix,
otherwise we try with a shorter one. 

The idea of a binary search for a matching prefix 
is the basic principle of
the van Emde Boas tree~\cite{vB77,vBKZ77,WillLog}.
However, a van Emde Boas tree is not just a plain binary trie
represented as above. One problem is the space requirements; 
a plain binary trie storing $d$ keys may contain as much as $\Theta(d\wordlength)$ nodes.
In a van Emde Boas tree, the number of nodes is decreased to $O(d)$ by
careful optimization.

In our application $\Theta(d\wordlength)$ nodes can be allowed. Therefore,
to keep things simple,
we use a plain binary trie.

\begin{lemma}
\label{lem:vEBgen}
A static data structure containing $d$ keys
and supporting neighbor queries 
in~$O(\log \wordlength)$ worst-case time
can be constructed in $O\paren{d^4}$ time and space.
The implementation can be done without division.
\end{lemma}

\begin{proof}
We study two cases.

Case 1: \hbox{$\wordlength > d^{1/3}$.} Lemma \ref{lem:FWgen} gives constant query cost.

Case 2: \hbox{$\wordlength \leq d^{1/3}$.}
In $O(d\wordlength)= o(d^2)$ time and space we construct a binary trie  
of height~$\wordlength$ containing all $d$ keys. 
Each key is stored at the bottom of a path of length $\wordlength$ and the keys
are linked together. In order to support
neighbor queries, each unary node 
contains a neighbor pointer to the next (or previous) leaf
according to the inorder traversal.

To allow fast access to
an arbitrary node, we store all nodes in a perfect hash table such that
each node of depth $i$ is represented by the $i$ bits on the path down to the node.
Since the paths are of different length, we 
use $\wordlength$ hash tables, one for each path length. Each hash table contains 
at most~$d$ nodes.
The algorithm by Fredman, Komlos, and
Szemeredi~\cite{FredStor} constructs a hash table of $d$ keys in $O(d^3 \wordlength)$ time.
The algorithm uses division, this can be avoided by
simulating each division in $O(\wordlength)$ time.
With this extra cost,
and since we use $\wordlength$ tables, the total construction time is $O\paren{d^3 \wordlength^3}= O(d^4)$
while the space is $O(d\wordlength)= o(d^2)$.

With this data structure, we can search for a key $x$ in $O(\log \wordlength)$ time by 
a binary search for the
node corresponding to $x$'s longest matching prefix. 
This search either ends at the bottom of the trie or at a unary node,
from which we find the closest neighboring leaf by following the 
node's neighbor pointer.

During a search, evaluation of the hash function
requires integer division. However, as pointed
out by Knuth~\cite{Kn3}, division with some precomputed constant $p$ may
essentially be replaced by multiplication with $1/p$.
Having computed $r= \floor{2^\wordlength/p}$ once in $O(\wordlength)$ time,
we can compute $x \ \idiv\ p$ as $\floor{xr / 2^\wordlength}$
where the last division is just a right shift $\wordlength$ positions. 
Since $\floor{x/p}-1 < \floor{xr / 2^\wordlength} \leq \floor{x/p}$ we can compute the
correct value of $x \ \idiv\ p$ by an additional test.
Once we can compute \idiv, we can also compute \mod.
\end{proof}

An alternative method for perfect hashing without division is the one
recently developed by Raman~\cite{Ram96}. Not only does this
algorithm avoid division, it is also asymptotically faster, $O(d^2 \wordlength)$.

\begin{corollary}
\label{cor:vEBpart}
There is a data structure occupying linear
space for which the
worst-case cost of a search and the amortized
cost of an update is
$
O\paren{\log \wordlength\log\log n}.
$
\end{corollary}
\begin{proof}
Let $T(n)$ be the worst-case search cost.
Combining Lemmas \ref{thm:search-red} and \ref{lem:vEBgen}
gives 
$
T(n) = O\paren{\log \wordlength} + T\paren{n^{4/5}}. 
$
\end{proof}

\subsection{Finishing the proof of Theorem \ref{thm:static-search}}

\noindent
If we combine Lemmas \ref{lem:FWgen} 
and \ref{lem:vEBgen}, we can in polynomial time construct
a dictionary over $d$ keys supporting searches in
time $S(d)$, where
\begin{equation}
\label{eq:totrec}
S(n) = O\paren{\min\paren{
% \begin{array}{l}
1+\frac{\log n}{\log \wordlength}, \log \wordlength
% \end{array}
}}
\end{equation}
Furthermore, balancing the two parts of the min-expression gives
$$
S(n) = O\paren{\sqrt{\log n}}.
$$
To get AC$^0$ bound in Theorem \ref{thm:static-search}, we 
combine some known results. From Andersson's packed B-trees \cite{And95},
it follows that if in polynomial time and space, we build a static 
AC$^0$ dictionary with membership queries in time $t$, then in 
polynomial time and space, we can build a static search structure
with operation time $O(\min_i\{it+\log n/i\})$. In addition, 
Brodnik et.al. \cite{BMM97} have
shown that such a static dictionary, using only standard AC$^0$ operations,
can be built with membership queries in time $t=O((\log n)^{1/2+o(1)})$.
We get the desired static search time by setting 
$i=O((\log n)^{1/4+o(1)})$. {\bf This completes the
proof of Theorem \ref{thm:static-search}, hence of Corollary \ref{cor:search}}.

\subsection{Two additional notes on searching}

\noindent
Firstly, we give the first 
deterministic polynomial-time (in $n$)
algorithm for constructing a linear space
static dictionary with $O(1)$ worst-case
access cost (cf. perfect hashing).

As mentioned earlier, a linear space data structure that 
supports member queries
(neighbor queries are not supported) 
in constant time can be constructed at a worst-case cost $O\paren{n^2 \wordlength}$
without division~\cite{Ram96}. 
We show that the dependency of word size can be removed.

\begin{proposition}
\label{obs:hash}
A linear space static data structure supporting member queries at a worst
case cost of $O(1)$ can be constructed 
in $O\paren{n^{2+\epsilon}}$ worst-case time.
Both construction and
searching can be done without division.
\end{proposition}

\begin{proof}
W.l.o.g we assume that $\epsilon < 1/6$.

Since Raman has shown that a perfect hash
function can be constructed in $O\paren{n^2 \wordlength}$ time without 
division)~\cite{Ram96}, we are done
for $n \geq \wordlength^{1/\epsilon}$.

If, on the other hand, $n < \wordlength^{1/\epsilon}$, we construct a static tree
of fusion tree nodes with degree $O\paren{n^{1/3}}$.
This degree is possible since $\epsilon < 1/6$.
The height of this tree is $O(1)$, the cost of constructing a node
is $O\paren{n^{4/3}}$ and the total number of nodes is $O\paren{n^{2/3}}$.
Thus, the total construction cost for the tree is $O\paren{n^2}$.

It remains to show that the space taken by the fusion tree nodes
is $O(n)$.
According to Fredman and Willard, a fusion tree node
of degree $d$ requires $\Theta\paren{d^2}$ space. This space 
is occupied by a lookup table where each entry contains a 
rank between $0$ and $d$. 
A space of $\Theta\paren{d^2}$ is small enough for the original fusion tree
as well as for our exponential search tree. However, in order to
prove this proposition, we need to reduce the space taken by a
fusion tree node from $\Theta\paren{d^2}$ to $\Theta\paren{d}$.
Fortunately, this reduction is straightforward. 
We note that a number between $0$ and $d$
can be stored in $\log d$ bits. Thus, since $d < \wordlength^{1/6}$, the total 
number of bits occupied by the lookup table 
is $O\paren{d^2 \log d} = O(\wordlength)$. This packing of numbers is done
cost efficiently by standard techniques. 

We conclude that
instead of $\Theta\paren{d^2}$, 
the space taken by the lookup table in a fusion tree node
is $O(1)$ ($O(d)$ would have been good enough).
Therefore, the space occupied by a fusion tree node can be made linear in
its degree.
\end{proof}

Secondly, we show how to adapt our data structure 
to certain input distribution.

In some applications, we may assume that
the input distribution is favorable. These kind
of assumptions may lead to a number of heuristic
algorithms and data structures whose analysis
are based on probabilistic methods. Typically,
the input keys may be assumed to be generated
as independent stochastic variables from some
(known or unknown) distribution; the goal is
to find an algorithm with a good expected behavior.
For these purposes, a deterministic algorithm is not
needed.

However, instead of modeling input as the result
of a stochastic process, we may characterize its
properties in terms of a {\em measure}. 
Attention is then moved from the process of generating
data to the properties of the data itself. In this
context, it makes sense to use a deterministic 
algorithm; given the value of a certain measure
the algorithm has a guaranteed cost.

We give one example of how to adapt our data structure according
to a natural measure. 
An indication of how ``hard'' it is to search for a key
is how large part of it must be read in order to distinguish it from
the other keys. We say that this part is the key's {\em distinguishing
prefix}. (In Section \ref{sec:trie} we used the term longest
matching prefix for essentially the same entity.)
For $\wordlength$-bit keys, the longest possible distinguishing prefix
is of length $\wordlength$. 
Typically, if the input is nicely distributed,
the average length of the distinguishing prefixes is $O(\log n)$.

As stated in Proposition~\ref{obs:adapt}, we can search faster when a
key has a short distinguishing prefix.

\begin{proposition}
\label{obs:adapt}
There exist a linear-space 
data structure for which the worst-case cost of a search and
the amortized cost of an update is
$O(\log b \log\log n)$ where $b \leq \wordlength$ is the length of the query key's
distinguishing prefix, i.e. the prefix that needs to be inspected
in order to distinguish it from each of the other stored keys.
\end{proposition}

\begin{proof}
We use exactly the same data structure as in
Corollary~\ref{cor:vEBpart}, with the same restructuring cost of
$O(\log\log n)$ per update.  The only difference is that we change the
search algorithm from the proof of Lemma~\ref{lem:vEBgen}.  Applying
an idea of Chen and Reif~\cite{ChenUsin}, we replace the binary search
for the longest matching (distinguishing) prefix by an
exponential-and-binary search. Then, at each node in the exponential
search tree, the search cost will decrease from $O(\log \wordlength)$
to $O(\log b)$ for a key with a distinguishing prefix of length~$b$.
\end{proof}

\section{Finger search and finger updates}
\label{sec:finger}
Recall that we have a finger pointing at a key $x$ while searching for 
another key $y$, and let $q$ be the number of keys between $x$ and $y$.
W.l.o.g. we assume $y>x$. In its traditional formulation,
the idea of finger search is that we should
be able to find $y$ quickly if $q$ is small. 
Here, we also consider
another possibility: the search should be fast if $y-x$ is small.
Compared with the data structure for plain searching, we need
some modifications to support finger search and updates efficiently.
The overall goal of this section is to prove
the statement of 
Theorem~\ref{thm:finger}:
\begin{quote}\it
There is a fully-dynamic deterministic linear space search structure that
supports finger updates in constant time, and
given a finger to a stored key $x$, searches
a key $y>x$ in time
$$
O\left(\min\left\{
\begin{array}{l} 
\sqrt{\log q/\log\log q} 
\\
\log \log q\cdot \frac{\log \log (y-x)}{\log\log\log (y-x)}
\\ 
\log\log q + \frac{\log q}{\log \wordlength} 
\end{array}
\right\}\right)$$
where $q$ is the number of stored keys between $x$ and $y$. If we
restrict ourselves to AC$^0$ operations, we still get a bound
of $O((\log q)^{3/4+o(1)})$.
\end{quote}
Below, we will first show how to reduce the cost of finger updates 
from the $O(\log\log n)$ in the last section to a constant. This
will then be combined with efficient static finger search structures.

\subsection{Constant time finger update}
In this section, we will generally show how to reduce the finger
update time of $O(\log\log n)$ from Lemma~\ref{lem:simple-update} to
a constant. The $O(\log\log n)$ bound stems from the fact when we insert
or delete a leaf, we use 
a local update step for each level above the leaf. Now, however, we
only want to use a constant number of local update steps in connection
with each leaf update. The price is that we have
less local update steps available for the postprocessing of join and splits. 
More precisely, we will prove the following
analogue to the general balancing in Theorem
\ref{thm:balance}:
\begin{theorem}\label{thm:finger-balance} 
Given a number series $n_1, n_2, \ldots$, with $n_1\geq 84$, $18 n_i< n_{i+1} <
n_i^2$ for $i\geq 1$, 
we can schedule split and joins to maintain a multiway tree where
each non-root node $v$ on height $i>0$ has weight at between $n_i/4$ and $n_i$.
A root node on height $h>0$ has weight at most $n_h$ and at least
$2$ children. The schedule gives the following properties:

(i) When a leaf $v$ is inserted or deleted, the schedule
uses a constant number of local update steps. The additional
time used by the schedule is constant.

(ii) For each split or join at level $i>1$
the schedule ensures that we have at least $\sqrt{n_i}$ local update steps 
available for postprocessing, including one in connection with
the split or join itself. For level $1$, we have $n_1$ local updates
for the postprocessing.
\end{theorem}
As we shall see later, the $\sqrt{n_i}$ local update steps suffice for
the maintenance of $S$-structures. As for the Theorem \ref{thm:balance},
we have 
\begin{lemma}\label{lem:finger-parents} In Theorem \ref{thm:finger-balance}, 
the parent of any node can be computed in constant time. 
\end{lemma}
\begin{proof} We use exactly the same construction as for 
Lemma \ref{lem:parents}.  The critical point is that we for
the postprocessing have a number of updates which is the proportional
to the number of children of a node. This is trivially the case for
level $1$, and for higher levels $i$, the number of children is
at most $n_i/(n_{i-1}/4)=O(\sqrt{n_i})$.
\end{proof}
As in the proof of Theorem \ref{thm:balance}, we will actually use
the parents of Lemma \ref{lem:finger-parents} in the proof of
Theorem \ref{thm:finger-balance}. As argued at the end of
Section \ref{ssec:apply-prot} this does not lead to a circularity.

Abstractly, we will use the same schedule for join and splits 
as in the proof of Theorem~\ref{thm:balance}. However, we will not
perform as many local update steps during a join or split process.
Moreover, the structural implementation of a join or split will
await its first local update. 

We note that level $1$ is exceptional, in that we need
$n_1$ local updates for the split and join postprocessing. This is
trivially obtained if we with each leaf update make $84$ local
updates on any postprocessing involving or tied to the parent. For
any other level $i>1$, we need $\sqrt{n_i}$ local updates, which is
what is obtained below.

The result will be achieved by a
combination of techniques. We will use a tabulation technique for the
lower levels of the exponential search tree, and a scheduling idea of
Levcopoulos and Overmars \cite{LevcBala} for the upper levels.

\subsubsection{Constant update cost for small trees on 
the lower levels}\label{S:low-level}

In this subsection, we will consider small trees induced by lower
levels of the multiway tree from Theorem \ref{thm:finger-balance}.

One possibility for obtaining constant update cost for search
structures containing a few keys would have been to use atomic heaps
\cite{FW94}.  However, here we aim at a solution using only~AC$^0$
operations.  We will use tabulation. A tabulation technique for finger
updates was also used by Dietz and Raman \cite{DR94}. They achieved
constant finger update time and $O(\log q)$ finger search time, for
$q$ intermediate keys, in the comparison based model of computation.
However, their approach has a lower bound of $\Omega(\log q/\log\log
q)$ as it involves ranks \cite{FS89}, and would prevent us from
obtaining our $O(\sqrt{\log q/\log\log q})$ bound. Finally, we note
that our target is the general schedule for multiway trees in Theorem
\ref{thm:finger-balance} which is not restricted to search
applications. 

Below, we present a schedule satisfying the conditions
of Theorem \ref{thm:finger-balance} except that we need tables for an
efficient implementation.

Every time we insert or delete a leaf $u$, we will do $1000$ local update
steps from $u$. The place for these local update steps is determined based on a
system of marking and unmarking nodes.  To place a local update from
a leaf $u$, we find its nearest unmarked ancestor $v$. 
We then unmark all nodes on the path from $u$ to $v$ and mark $v$.
If $v$ is
involved in or tied to a join or split process, we perform one local
update step on this process. If not, we check if the weight of $v$ is
such that it should split or join or tie to a neighboring split or
join, as described in the schedule for Proposition \ref{lem:buckets}.
We then mark the involved nodes and perform a local update step at
each of them.
\begin{lemma} For a split or join process on level $i$, we get
at least $n_i/2^i$ local updates steps.
\end{lemma}
Note that $n_i>18^i$, so $n_i/2^i>\sqrt{n_i}$, as in Theorem
\ref{thm:finger-balance}.
\begin{proof}
  First, using a potential function argument, we analyze how
  many time a level $i$ node $v$ gets marked during $p$ local
  updates from leaves below $v$.
  The potential of a marked node is 0 while the potential of an
  unmarked node on level $i$ is $2^i$.  The sub-potential of $v$ is
  the sum of the potential of all nodes descending from or equal to
  $v$. Then, if an update below $v$ does not unmark $v$, it decreases the
  sub-potential by 1. On the other hand, if we unmark $v$, we also
  unmark all nodes on a path from a leaf to $v$, so we increase the
  potential by $2^{i+1}-1$. When nodes are joined and split, the
  involved nodes are all marked so the potential is not increased. The
  potential is always non-negative.  Further, its maximum is achieved
  if all nodes are unmarked. Since all nodes have degree at least 4,
  if all nodes are unmarked, the leaves carries for more than
  half of the potential.  On the other hand, the number of leaves
  below $v$ is at most $n_i$, so the maximum potential
  is less than $2n_i$. It follows that the number of times $v$ gets
  unmarked is more than $(p-2n_i)/2^{i+1}$. Hence $v$ gets
  marked at least $(p-2n_i)/2^{i+1}$ times.
  
  From Theorem \ref{thm:balance} we know that during a split or join
  process, there are at least $n_i/84$ leaf updates below the at most
  $4$ level $i$ nodes $v_0,v_1,v_2,v_3$ involved in or tied to the
  process. Each of these leaf updates results in $1000$ local
  updates. Thus, if $p_j$ is the number of leaf updates from
  below $v_j$ during our process, $p_0+p_1+p_2+p_3\geq 10n_i$.
  Consequently, the number of local updates for our process
  is 
\[\sum_{j=0}^3 (p_j-2n_i)/2^{i+1}\geq (10n_i-8n_i)/2^{i+1}=n_i/2^i,\]
  as desired.
\end{proof}
The above schedule, with the marking and unmarking of nodes to
determine the local update steps, could easily be implemented in time
proportional to the height of the tree, which is $O(\log\log n)$. To
get down to constant time, we will use tables of size $o(n)$ to deal
with small trees with up to $m$ nodes where $m=O(\sqrt{\log n})$. Here
we think of $n$ as a fixed capacity for the total number of
stored keys. As the number of actual keys change by a factor of $2$, we
can build a data structure with new capacity in the background.

Consider an exponential search tree $E$ with at most $m$ nodes.  With
every node, we are going to associate a unique index below $m$, which is given
to the node when it is first created by a split. Indices are recycled
when nodes disappear in connection with joins. We will have a table
of size $m$ that maps indices into the nodes in $E$. Conversely, 
with each node in $E$, we will store its index. In
connection with an update, tables will help us find the index to the
node to be marked, and then the table give us the corresponding
node. 

Together with the tree $E$, we store a bitstring $\tau_E$ representing
the topology of $E$. More precisely, $\tau_E$ represents the depth first
search traversal of $E$ where 1 means go down and 0 means go up. Hence,
$\tau_E$ has length $2m-2$. Moreover, we have a table $\mu_E$ that maps
depth first search numbers of nodes into indices. Also, we have a table
$\gamma_E$ that for every node tells if it is marked. 
We call 
$\alpha_E=(\tau_E,\mu_E,\gamma_E)$ the signature 
of $E$.  Note that we have $\leq 2^{2m}\times m^m\times 2^m\times O(m) =
m^{O(m)}$ different signatures.

For each of the signatures, we tabulate what to do in connection with
each possible leaf update. More precisely, for a leaf delete, we have
a table that takes a signature of tree and the index of the leaf to be
deleted and produces the signature of the tree without the leaf. Thus
when deleting a leaf, we first find its associated index so that we
can use the table to look up the new signature.  Similarly, for a leaf
insert, we assume the index of a preceding sibling, or the parent if
the leaf is to be a first child. The table should produce not only the
new signature, but also the index of the new leaf. This index should
be stored with the new leaf. Also, the leaf should be stored with the
index in the table mapping indices to nodes.

For the local updates, we have a table taking a signature and the index
of a leaf to do the local update from. The table produces the
index of the node to be marked, hence at which to do a local update. If a
split or join is to be done, the table tells the indices of involved
nodes. For a split, this includes the child at which to split the children
sequence. Also, it includes the index of the new node. Finally,
the table produces the signature of the resulting tree.
All the above mentioned tables can easily be constructed in $m^{O(m)}=o(n)$ 
time and space.

Let $a$ be such that $n_a\leq \sqrt{\log n}< n_{a+1}$ \mtch and set 
$m=n_a$. We are
going to use the above tabulation to deal with levels $0,...,a$ of the
multiway tree of Theorem \ref{thm:finger-balance}.  Note that if
$n_1>\sqrt{\log n}$, $a=0$, and then we can skip to the next
subsection (\S \ref{S:high-level}).
With each of the
level $a$ nodes, we store the signature of the descending subtree as
well as the table mapping indices to nodes. Also, with each leaf, we
store an ancestor pointer to its level $a$ ancestor. Then, when a leaf
is added, it copies the ancestor pointer of one of its siblings. Via
these ancestor pointers, we get the signature of the tree that is
being updated.

A new issue that arises is when level $a$ nodes $u$ and $v$ get joined 
into $u$. For this case, we temporarily allow indices up to
$2m-1$, and add $m$ to the indices of nodes descending from $v$.
A table takes the signatures of the subtrees of $u$ and $v$ and produce
the signature for the joined tree with these new indices. Also,
we place a forward pointer from $v$ to $u$, so that nodes
below $v$ can find their new ancestor in constant time. To get
the index of a node, we take its current ancestor pointer. If it 
points to a node with a forward pointer, we add $m$ to the 
stored index. Conversely, given an index, if it is not less than
$m$, this tells us that we should use the old table from $v$, though
subtracting $m$ from the index.

During the postprocessing of the join, we will traverse the subtree
that descended from $v$. We move each node $w$ to $u$, redirecting the
ancestor pointers to $u$ and give $w$ a new unique index below $m$.
Such an index exists because the total size of the tree after the join
is at most $m$.  The indexing is done using a table that suggests the
index and the resulting new signature.  The node is then inserted in the
table at $u$ mapping indices below $m$ to nodes. Since we use the same
general schedule as that in Theorem \ref{thm:balance}, we know that we
have $n_a/84$ updates below the join before the join needs to be
completed. In that time, we can make a post traversal of all the at
most $n_a$ descendants of the join, assigning new indices and updating
parent pointers. We only deal with a constant number of descendants at
the time. For the traversal, we can use a depth first traversal,
implemented locally as follows. At each point in time, we are at some
node $w$, going up or down. We start going down at that first child of
$v$ from when the join was made. If we are going down, we move $w$ to
its first child. If we are going up and there is a sibling to the
left, we go to that sibling, going down from there. If we are going up
and there is no sibling to the left, we go up to the parent. At each
node, we check if it has already been moved to $u$ by checking if
the ancestor pointer points to $u$. If we are
about to join or split the traversal node $w$, we first move $w$ away
a constant number of steps in the above traversal. This takes constant
time, and does not affect the time bounds for join and split.

A level $a$ split of $u$ into $u$ and $v$ is essentially symmetric
but simpler in that we do not need to change the indices. In the
traversal of the new subtree under $v$, we only need to redirect
the ancestor pointers to $v$ and to build the table mapping indices
to nodes in the new subtree. 

The traversals take constant time for level $a$ join and split processes 
for each descending leaf updates. In the next subsection, we are going
to do corresponding traversals for two other distinguished levels.

Including the new tables for index pairs, all tables are constructed
in $m^{O(m)}=o(n)$ time and space. With them, we implement the
schedule of Theorem \ref{thm:finger-balance} for levels $i=0,..,a$
using constant time and a constant number of local update steps per
leaf update, yet providing at least $\sqrt{n_i}$ local updates for the
postprocessing of each join or split.

\subsubsection{Moving up the levels}\label{S:high-level}
We are now going to implement the schedule of
Theorem~\ref{thm:finger-balance} on levels $a+1$ and above. \mtch In
connection with a leaf update, we have constant time access to its
level $a$ ancestor, hence also to its level $a+1$ ancestor.  We note
that if $n_1>\sqrt{\log n}$, $a=0$, and then we are not using any of
the material from the previous subsection (\S \ref{S:low-level}). Then
the whole construction will be implementable on a pointer machine.

To get to levels $a+1$ and above, we are going to use the following 
variant of a lemma of Overmars and 
Levcopoulos \cite{LevcBala}:
\begin{lemma}\label{lem:counters} 
Given $p$ counters, all starting at zero, and an adversary incrementing these
counters arbitrarily. Every time 
the adversary has made $q$ increments, the increments being by one at
the time, we subtract $q$ from some largest counter, or set
it to zero if it is below $q$. Then the largest possible
counter value is $\Theta(q\log p)$. 
\end{lemma}
In the original lemma from  \cite{LevcBala}, instead of subtracting
$q$ from a largest counter, they split it into
two counters of equal size. That does not imply our case, so we need our own 
proof, which also happens to be much shorter. 
\begin{proof}
We want to show that the maximal number of counters larger 
than $2iq$ is at most $p/2^i$. The proof is by induction. 
Obviously, the statement is true for $i=0$, so consider $i>0$.
Consider a time $t$ where the number of counters larger than
$p/2^i$ is maximized, and let $t^-$ be the last time before
$t$ at which the largest counter was $(2i-1)q$. 

We consider it one step to add $1$ to $q$ counters, and subtract
$q$ from a largest counter. Obviously, at the end of the day, we
can at most do $q$ better in total.

The basic observation is that between $t^-$ and $t$, no change
can increase the sum of the counter excesses above $(2i-2)q$, for
whenever we subtract $q$ it is from a counter which is
above $(2i-1)q$. However, at time $t^-$, by induction, we
had only $p/2^{i-1}$ counters above $(2i-2)q$, and each
had an excess of at most $q$. To get to $2iq$, a counter
needs twice this excess, and since the total excess can
only go down, this can happen for at most half the counters.
\end{proof}
For the implementation of Lemma \ref{lem:counters}, we have
\begin{lemma}\label{lem:impl-counter}
Spending constant time per counter increment in Lemma \ref{lem:counters},
the largest counter to be reduced can be found in constant time.
\end{lemma}
\begin{proof} We simply maintain a doubly linked sorted list of counter
values, and with each value we have a bucket with the counters with
that value. When a counter $c$ is increased from $x$ to $x+1$, we check
the value $x'$ after $x$ in the value list. If $x'>x+1$, we insert $x+1$
into the value list with an associated bucket. We know move $c$ to the
bucket of $x+1$, removing $x$ if its bucket gets empty. Decrements by one
can be handled symmetrically. Thus, when a largest counter $a$ has been
picked, during the next $k$ increments, we can decrement $a$ by one.
\end{proof}
We are going to use the above lemmas in two bands, one on levels
$a+1,...,b$ where $b$ is such that $n_b\leq (\log n)^{\log\log n}<
n_{b+1}$, and one levels $b+1$ and up. First, we consider levels
$a+1,...,b$.

To describe the basic idea, for simplicity, we temporarily assume that
there are no joins or splits.  Set $q=b-a$.
For $i=a+1,...,b$, during $\Omega(n_i)$ leaf updates
below a node $v$ on level $i$, we will get $\Omega(n_i/q)$ local updates
at $v$. Since $n_{i+1}>18 n_i$, $q<\log_{18} (n_b/n_a)<(\log\log n)^2$.
On the other hand, $n_i\geq n_{a+1}>\sqrt{\log n}$, so $q=o(\sqrt{n_i})$.

Each level $a+1$ node $v$ has a counter that is incremented every time
we have a leaf update below $v$.  In the degenerate case where $a=0$,
we always make a local update at $v$ so as to get enough updates on
level $1$ as required by Theorem~\ref{thm:finger-balance}. \mtch We
make an independent schedule for the subtree descending from each
level $b$ node $u$.  Once for every $q$ updates below $u$, we pick a
descending level $i$ node with the largest counter, do a local update
at $v$, and subtract $q$ from the counter. During the next $q-1$ leaf
updates below $u$, we follow the path up from $v$ to $u$, doing a
local update at each node on the way.

A largest counter below $u$ is maintained as described in 
Lemma~\ref{lem:impl-counter}. The number of counters below $u$ is
at most $p=n_b/(n_a/4)$, so by Lemma~\ref{lem:counters}, the
maximal counter value is $O(q\log p)=O((\log n_b)^2)=O((\log\log n)^4)$.

Now, for $i=a+1,...,b$, consider a level $i$ node $w$. The maximal
number of counters below $w$ is $n_i/(4n_{a+1})$, so their total
value is at most 
\[O((n_i/n_{a+1})(\log\log n)^4)=O((n_i/\sqrt{\log n})(\log\log n)^4)=o(n_i).\]
Each update below $w$ adds one to this number. Moreover, we
do a local update at $w$ every time we subtract $q$ from one of the
descending counters, possibly excluding the very last subtraction if we have
not passed $w$ on the path up to $u$. Consequently, during $r=\Omega(n_i)$
leaf updates below $w$, the number of local updates at $w$ is at least
\[(r-o(n_i)-q)/q=\Omega(n_i/q)=\omega(\sqrt{n_i}).\]
Next, we show how to maintain approximate weights. For the nodes $v$ on level $a+1$, we assume we
know the exact weight $W_v$. For  nodes $w$ on
levels $i=a+1,...,b$, we have an approximate weight $\widehat W_w$.
When the counter of a level $a+1$ node $v$ is picked, we set
$\Delta=W_v-\widehat W_v$ and $\widehat W_v=W_v$. As we
move up from $v$ to $u$ during the next $q-1$ updates, at each
node $w$, we set $\widehat W_v=\widehat W_v+\Delta$.

We will now argue that for any node $w$ on level $i=a+1,...,b$, the
absolute error in our approximate weight $\widehat W_w$ is $o(n_i)$. 
The error  in $\widehat W_w$ is at most the sum of the counters below $w$ 
plus $q$, and we have already seen above that this value is $o(n_i)$. It
follows that 
\[W_w=(1\pm o(1))\widehat W_w.\]
This error is small enough that we can use the approximate weights for
the scheduling of split and joins. More precisely, in the analysis,
we rounded at various points, and the rounding left room for errors below
a constant fraction.

We are now ready to describe the details of the schedule as nodes get
joined and split. From the last subsection, we have ancestor pointers
to level $a$, and via table we can also get the exact weight. From
this, we can easily get ancestor pointers and exact weights on level
$a+1$. On level $a+1$, we can then run the join and split schedule from
Section \ref{ssec:apply-prot}. 

For level $i=a+2,...,b$, we use the approximate weights both for
the nodes and for the children. When we get a local update at
a node $w$, we know that $\widehat W_w$ has just been updated
and that it equals the sum of the weights of the children, so
we do have local consistency in the approximate weights. We then
use the new approximate weight in the schedule of Proposition
\ref{lem:buckets} to check if $w$ is to be joined or split or
tied to some other join or split process. The local update
step is applied to any join or split process neighboring $w$.

Finally, we use the traversal technique from the last subsection
to maintain ancestor pointers to level $b$ nodes. This means
that we use constant time on level $b$ in connection with each
leaf update. In connection with a join or split postprocessing
on level $b$, this time also suffice to join or split the priority
queue over counters below the processed nodes. This completes
our maintenance of levels $a+1,....,b$.

For the levels above $b$, we use the same technique as we did for
levels $a+1,....,b$, but with the simplification that we have only one
tree induced by levels above $b$. Consequently, we have only one
priority queue over all counters on level $b$.  The numbers, however,
are a bit different.  This time, the number $q'$ of levels is
$\log_{18} (n/n_b)<\log n$.  However, for $i>b$, $n_i>(\log
n)^{\log\log n}$, so $q'=o(\sqrt{n_i})$.

We have one priority queue over all counters on level $b$, of which
there are at most $p'=n/(n_{b+1}/4)$, so by Lemma~\ref{lem:counters}, the
maximal counter value is $O(q'\log p')=O(\log n(\log\log n)^2)$.

Now, for $i>b$, consider a level $i$ node $w$. The maximal
number of counters below $w$ is $n_i/(4n_{b+1})$, so their total
value is at most 
\[O((n_i/n_{b+1})\log n(\log\log n)^2)=O((n_i/\log n^{\log\log n})
\log n(\log\log n)^2)=o(n_i).\]
With the above changes in numbers, we use the same technique for
levels above $b$ as we used for levels $a+1,...,b$. 
{\bf This completes the proof of Theorem \ref{thm:finger-balance}}

\begin{corollary}\label{cor:finger-search-balance}
Given a number series $n_0,n_1, n_2<, \ldots$, with $n_0=1$, $n_1\geq 84$, 
$n_i^2>n_{i+1} >
18 n_i$, we maintain a multiway tree where each node on height $i$ 
which is neither the root nor a leaf node has weight between $n_i/4$ and 
$n_i$. If an $S$-structure for a node on height $i$ 
can be built in $O(\sqrt{n_{i-1}})$ time, or $O(n_1)$ time for
level $1$, we can maintain $S$-structures
for the whole tree in constant time per finger update.
\end{corollary}
\begin{proof} We use the same proof as the one we used to
prove Corollary \ref{cor:search-balance} from Theorem \ref{thm:finger-balance}.
\end{proof}

\subsection{Fast finger search}

We first concentrate on implementing a fast finger search, postponing
the constant time finger updates to the next subsections.  For
simplicity, we will always assume that the fingered key $x$ is smaller
than the key $y$ sought for. The other case can be handled by a
``mirror'' data structure where each key $x$ is replaced by $U-x$,
where $U$ is the largest possible key.

The following finger search analog of Theorem 
\ref{thm:search-red} is obtained using the same kind of methods
as for pointer based finger search structures, i.e. by the use of 
horizontal links.
\begin{theorem}\label{lem:finger-red-top}
Suppose a static search structure on
$d$ integer keys can be constructed in $O(d^{(k-1)/2})$, $k\geq 2$,  
time and space so given a finger to a stored key $x$, we can search
a key $y>x$ in time $S(d,y-x)$. We can then construct
a dynamic linear space search structure that with $n$ integer keys supports 
finger updates in time constant time and 
finger searches in time $T(q,y-x)$ where $q$ is the number of
stored keys between $x$ and $y$ and
$T(n)\leq T(n^{1-1/k})+O(S(n,y-x))$. Here $S$ is supposed
to be non-decreasing in both arguments. The reduction itself uses only 
standard AC$^0$ operations.
\end{theorem}
\begin{proof}
We use an exponential search tree where on each level we have horizontal
links between neighboring nodes. It is trivial to modify join and
split to leave horizontal pointers between neighboring nodes on the
same level. 

A level $i$ node has $O(n_i/n_{i-1})=O(n_i^{1/k})$ children, so, by
assumption, its $S$-structure is built in
time $O(n_i^{(k-1)/(2k)})=O(\sqrt{n_{i-1}})$. Hence we can
apply Corollary \ref{cor:finger-search-balance}, and maintain
$S$-structures at all nodes in constant time per finger update.

To search for $y>x$, given a finger to $x$, 
we first traverse the path up the tree from the leaf containing $x$. 
At each level, we examine
the current node and its right neighbor until a node $v$ is found
that contains $y$. Here the right neighbor is found in constant time
using the horizontal links between neighbors. As we shall see later, 
the node $v$ has the advantage that its largest possible
degree is closely related to $q$.

Let $u$ be the child of $v$ containing $x$ and let $x'$ be
the separator immediately to the right of $u$.
Then, $x\leq x'\leq y$, and if we start our search from $x'$,
we will find the child $w$ where $y$ belongs in $S(d,y-x') \leq S(d,y-x)$ time,
where $d$ is the degree of $v$.

We now search down from $w$ for $y$. At 
each visited node, the left splitter $x'$ satisfies $x\leq x'\leq y$
so we start our search from the left splitter.

We are now going to argue that the search time is $T(q,y-x)\leq
T(q^{1-1/k})+O(S(q,y-x))$, as stated in the lemma. Let $i$ be the
level of the node $v$. Let $u$ be the level $i-1$ ancestor of the leaf
containing $x$, and let $u'$ be the right neighbor of $u$. By definition of
$v$, $y$ does not belong to $u'$, and hence all keys below $u'$ are
between $x$ and $y$. It follows that $q\geq n(u')\geq
n_{i-1}/10$. Now, the recursive search bound follows using the
argument from the proof of Lemma \ref{lem:search-red}.
\end{proof}

Note in the above lemma, that it does not matter whether the static 
search structure supports efficient finger search in terms of the number $d$ 
of intermediate keys. For example, the static search bound of 
$O(\sqrt{\log n/\log\log n})$ from \cite{BF99} immediately implies 
a dynamic finger search bound of $O(\sqrt{\log q/\log\log q})$ where
$q$ is the number of stored keys between the fingered key $x$ and the 
sought key $y$. However, if we want efficiency in terms of $y-x$, we
need the following result.
\begin{lemma}
\label{lem:loglogxy}
A data structure storing a set $X$ of $d$ keys from a universe of size $U$
can be constructed in $d^{O(1)}$ time and space such that
given a finger to stored key $x\in X$, we search a key $y>x$ in 
time~$O(\log\log (y-x)/ \log\log\log (y-x))$.
\end{lemma}
\begin{proof}
Beame and Fich \cite{BF99} have shown that a polynomial space
search structure can be constructed with search time 
$O(\min\{\sqrt{\log n/\log\log n},\log\log U/\log\log U\})$, where $n$ 
is the number of keys and $U = 2^{\wordlength}$ is the size of the universe they are drawn
from. As a start, we will have one such structure over our $d$ keys.
This gives us a search time of $O(\sqrt{\log d/\log\log d})$. Hence
we are done if 
 $\log\log (y-x)/ \log\log\log (y-x)=\Omega(\sqrt{\log d/\log\log d})$, and
this is the case if $y-x\geq 2^d$.

Now, for each key $x\in X$, and for $i=0,...,\log\log d$, we will
have a search structure $S_{x,i}$ over the keys in the range
$[x,x+2^{2^{2^i}})$, with search time $O(\log\log
2^{2^{2^i}}/\log\log\log 2^{2^{2^i}})=O(2^i/i)$. Then to find $y<x+2^d$,
we look in $S_{x,\lceil \log\log\log (y-x)\rceil}$. Now,
$2^{2^{2^{\lceil \log\log\log (y-x)\rceil}}}<(y-x)^{\log(y-x)}$, the
search time is $O(\log\log (y-x)^{\log(y-x)}/\log\log\log (y-x)^{\log(y-x)})=
O(\log\log (y-x)/\log\log\log (y-x))$.

It should be noted that it is not a problem to find the appropriate
$S_{x,i}$. Even if for each $x$, we store the $S_{x,i}$ as a linked
list together with the upper limit value of $x+2^{2^{2^i}}$, we can get
to the appropriate $S_{x,i}$ by starting at $S_{x,0}$ and moving
to larger $S_{x,i}$ until $y<x+2^{2^{2^i}}$. This 
takes $O(\log\log\log (y-x))=o(\log\log (y-x)/\log\log\log (y-x))$ steps.

Finally, concerning space and construction time, since we only have
$O(\log\log d)$ search structures for each of the $d$ elements in $X$,
polynomiality follows from polynomiality of the search structure of
Beame and Fich.
\end{proof}
\begin{prooft}{thm:finger} The result follows directly from 
the reduction of Theorem \ref{lem:finger-red-top} together with
the static search structures in Theorem \ref{thm:static-search},
Theorem \ref{thm:BF99}, and Lemma \ref{lem:loglogxy}.
\end{prooft}

\section{String searching}\label{sec:string}
In this section, we prove Theorem \ref{thm:string}.

\subsection{Preliminaries}
Our string searching result utilizes Corollary \ref{cor:search} and
Proposition \ref{obs:hash}.

\paragraph{Tries}
As a basic component, we use a trie over the strings where the characters are
1-word integers \cite[\S III]{Meh84}.  For technical reasons, we
assume that each string ends with a special character $\bot$, hence that no
string is a prefix of any other string. Abstractly, a trie over a set
$S$ of strings is the rooted tree whose nodes are the prefixes of
strings in $S$. For a string $\alpha$ and a 1-word character $a$,
the node $\alpha a$ 
has parent $\alpha$ and is labeled $a$.  The root is not labeled,
so $\alpha a$ is the labels encountered on the path from the root to
the node $\alpha a$. Our trie is ordered in the sense that the
children of a node are ordered according to their labels.  
We use standard path (or Patricia) compression, so
paths of unary nodes are stored implicitly by
pointers to the stored strings. Hence the trie data structure is
really concentrated on the $O(n)$ branching nodes. 

By storing appropriate
pointers, the problem of searching a string $x$ among the stored
strings $S$ reduces to (1) finding the longest common prefix
$\alpha$ between $x$ and the strings in $S$, and (2) searching the
next 1-word character of $x$ among the labels of the children of the
trie node $\alpha$. 
In a static implementation, we would use a dictionary in each node,
which would allow us to spend constant time at each visited node during 
step (1). Then, by 
keeping a search structure from Corollary \ref{cor:search} 
at each branching node, we perform step (2) in
$O(\sqrt{\log n/\log\log n})$ time, which is fine. 

However, in a dynamic setting we cannot use dictionaries
in each node over all children since we cannot update linear
spaced dictionaries
efficiently in the worst case. Instead, we will sometimes allow step (1)
to spend more than constant time in each visited node. This is fine
as long as the total time spent in step (1) does not exceed the total
bound aimed at.

\subsection{Efficient traversal down a trie}
Our new idea is to only apply the constant time dictionaries
to some of the children.
In a trie node, we differ between "heavy" and "light" children,
depending on their number of descending leaves. The point is that
heavy children will remain for a long time, and hence we can store them in
a dictionary which is rebuilt during a relatively slow
rebuilding process. For light children, we cannot use a dictionary, 
instead we store them in a dynamic search structure 
from Corollary \ref{cor:search}.
Although this will give rise to a non-constant search time, we are
still fine since the low weight of the found child will guarantee that
the problem size has decreased enough to compensate for the search effort.

In more detail: At a node with weight $m$, we only 
store heavy children with $\Omega(m^{1-1/k})$ descending keys
in a dictionary, 
where $k=2+\eps$ is the exponent from the dictionaries in 
Proposition \ref{obs:hash}, the other children are stored in a dynamic search 
structure (an exponential search tree). 
Our string searching time is then $O(\ell)$ 
for the use of dictionaries
and for following pointers. The total cost of using the search structures
bounded by $T(n)$, where
\begin{eqnarray*}
T(m)&\leq&O(\sqrt{\log m/\log\log m})+T(m^{1-1/k})\\
&=&O(\sqrt{\log m/\log\log m}).
\end{eqnarray*}
Adding up, our total time bound is $O(\sqrt{\log n/\log\log n}+\ell)$, which
is optimal.

We maintain the dictionary of a node $v$ by periodic rebuilding. We maintain
an unordered list of all children with more than $m^{1-1/k}/2$ descendants. 
In every period, we first scan the list, and then build a dictionary
over the labels of the scanned children. When the new dictionary is completed
it replaces the previous one in constant time.
There are only $O(m^{1/k})$ labels,
so this takes $O(m^{1/k\cdot(k-1)})=O(m^{1-1/k})$ time. Hence, spending
$O(1)$ time per update to a descendant, 
we can complete a period for every $m^{1-1/k}/4$ updates, and this 
ascertains that no child can contain more than $m^{1-1/k}$ children
without being in the current dictionary.

The space bound is proven in a rather straightforward manner.

\paragraph{Practical simplifications}
In our reduction, we used a polynomial spaced dictionary
from Proposition \ref{obs:hash}. By increasing the exponent, 
we can allow ourself
to use even simpler and faster hashing schemes, such as 1-level
hashing with quadratic space, which would remove collision handling.
This way of using space seems to be a good idea also for
more practical randomized data structures.

\section{Other applications of our techniques}\label{sec:other-appl}
In this section we discuss how the techniques presented in this
paper have been applied in other contexts.

Variants of exponential search trees have been instrumental in many
of the previous strongest results on deterministic linear integer
space sorting and priority queues \cite{And96,Tho98,AT00,Han01}. Here
a priority queue is a dynamic set for which we maintain the minimum
element.  When first introduced by Andersson \cite{And96}, they
provided the then strongest time bounds of $O(\sqrt{\log n})$ for
priority queues and $O(n\sqrt{\log n})$ for sorting.  As noted by
Thorup in \cite{Tho98}, we can surpass the $\Omega(\sqrt{\log
  d/\log\log d})$ lower bound for static polynomial space searching in
a set of size $d$ if instead of processing one search at the time, we
process a batch of $d$ searches. Thorup got the time per key in the
batch down to $O(\log\log d)$. In order to exploit this, Thorup
developed an exponential priority queue tree where the update time was
bounded by \req{eq:rec}, but with $S(n)$ being the per key cost of
batched searching. Thus he got priority queues with an update time of
$O((\log\log n)^2)$ and hence sorting in $O(n(\log\log n)^2)$ time.
Thorup's original construction was amortized, but a worst-case
construction was later presented by Andersson and Thorup \cite{AT00}.
More advanced static structures for batched searching where later
developed by Han \cite{Han01} who also increased the batch size to
$d^2$. He then ended up with a priority queue update time $O((\log\log
n)(\log\log\log n))$ and sorting in $O(n(\log\log n)(\log\log\log n))$
time.  However, exponential search trees are not used in Han's recent
deterministic $O(n\log\log n)$ time sorting in linear space
\cite{Han02} or in Thorup's \cite{Tho02} corresponding priority queue
with $O(\log\log n)$ update time.  Since \req{eq:rec} cannot give
bounds below $O(\log\log n)$ per key, so it looks as if the role of
exponential search trees is played out in the context of integer
sorting and priority queues.

Recently, Bender, Cole, and Raman \cite{BCR02} have used 
the techniques for to derive worst-case efficient cache-oblivious
algorithms for several data structure problem. This nicely highlights
that the exponential search trees themselves are not restricted to integer
domains. It just happens that our applications in this paper are for
integers.

Theorem \ref{thm:balance} provides a general tool for maintaining
balance in multiway trees. These kind of techniques have been used
before, but they have never been described in an such a general
independent quotable way.  By using our theorems, many proofs of
dynamization can be simplified, and in particular, we can avoid the
standard hand-waving, claiming without proper proof that amortized
constructions can be deamortized.  The second author \cite{Tho02} has
recently used our Proposition \ref{lem:buckets} in a general reduction
from priority queue to sorting, providing a priority queue whose
update cost is the per key cost of sorting. Also, he \cite{Tho02a} has
recently used Theorem \ref{thm:balance} in a space efficient solution
to dynamic stabbing, i.e., the problem of maintaining a dynamic set of
intervals where the query is to find an interval containing a given
point. This codes problems like method look-up in object oriented
programming and IP classification for firewalls on the internet. The
solution has query time $O(k)$, update time $O(n^{1/k})$, and uses
linear space. Previous solutions used space $O(n^{1+1/k})$. The
solution does not involve any search structure, so it is important
that Theorem \ref{thm:balance} has a general format not specialized to
search applications.

\section{An open problem}\label{sec:open-problem}
It is an interesting open problem what is the right complexity
for searching with standard, or even non-standard, AC$^0$ operations?
Andersson et.al. \cite{AMRT96}, have shown that even if we allow
non-standard AC$^0$ operations, the exact complexity of membership queries
is $\Theta(\sqrt{\log/\log\log n})$. This contrast the situation at the RAM,
where we can get down to constant time for membership queries. Interestingly,
$\Theta(\sqrt{\log/\log\log n})$ is also the RAM lower bound for
searching, so the question is potentially, it is possible to
do the $\Theta(\sqrt{\log/\log\log n})$ searching using AC$^0$ operations
only.

%\bibliographystyle{plain}
%\bibliography{paper}

\end{document}